\documentclass[a4paper, amsfonts, amssymb, amsmath, reprint, showkeys, nofootinbib, twoside]{revtex4-1}
\usepackage[english]{babel}
\usepackage[utf8]{inputenc}
\usepackage{appendix}

\usepackage[plain]{algorithm}
\usepackage{amsmath}
\usepackage{amssymb}
\usepackage{bm}
\usepackage[nolabel]{showlabels}
\usepackage[ISO=false]{diffcoeff}
\usepackage{listings}
\usepackage{graphicx}
\usepackage[utf8]{inputenc}
\usepackage{tablefootnote}
\usepackage{threeparttable}
\usepackage{makecell}
\usepackage{xcolor}
\definecolor{mydarkgray}{gray}{0.2}

\usepackage{graphics}

\renewcommand{\b}[1]{\mathbf{#1}}

\newcommand{\avg}[1]{\left\langle{#1}\right\rangle}

\renewcommand{\v}[1]{\mathbf{#1}}

\newcommand{\var}[1]{\operatorname{var}\left[#1\right]}

\newcommand{\qb}{\bar{q}_{\theta}}
\newcommand{\q}{q_{\theta}}
\newcommand{\whb}{\hat{\bar{w}}}

\usepackage{graphicx}
\usepackage{dcolumn}
\usepackage{bm}
\usepackage{textcomp}
\usepackage{color}
\usepackage{amsmath}
\usepackage[ISO=false]{diffcoeff}

\renewcommand{\b}[1]{\mathbf{#1}}

\begin{document}

\title{Simulating first-order phase transition with hierarchical autoregressive networks}

\author{Piotr Białas}
\email{piotr.bialas@uj.edu.pl}
\affiliation{Institute of Applied Computer Science, Jagiellonian University, ul. \L ojasiewicza 11, 30-348 Krak\'ow, Poland}
\author{Paulina Czarnota}
\email{p.czarnota@student.uj.edu.pl}
\affiliation{Faculty of Biochemistry, Biophysics and Biotechnology, Jagiellonian University, ul.~Gronostajowa 7, 30-387 Kraków, Poland}
\author{Piotr Korcyl}
\email{piotr.korcyl@uj.edu.pl}
\author{Tomasz Stebel}
\email{tomasz.stebel@uj.edu.pl}
\affiliation{Institute of Theoretical Physics, Jagiellonian University, ul. \L ojasiewicza 11, 30-348 Krak\'ow, Poland}

\date{\today}

\begin{abstract}
We apply the Hierarchical Autoregressive Neural (HAN) network sampling algorithm to the two-dimensional $Q$-state Potts model and perform simulations around the phase transition at $Q=12$. We quantify the performance of the approach in the vicinity of the first-order phase transition and compare it with that of the Wolff cluster algorithm. We find a significant improvement as far as the statistical uncertainty is concerned at a similar numerical effort. In order to efficiently train large neural networks we introduce the technique of pre-training. It allows to train some neural networks using smaller system sizes and then employing them as starting configurations for larger system sizes. This is possible due to the recursive construction of our hierarchical approach. 
Our results serve as a demonstration of the performance of the hierarchical approach for systems exhibiting bimodal distributions. Additionally, we provide estimates of the free energy and entropy in the vicinity of the phase transition with statistical uncertainties of the order of $10^{-7}$ for the former and $10^{-3}$ for the latter based on a statistics of $10^6$ configurations.
\end{abstract}


\maketitle

\section{Introduction}

Numerical Monte Carlo simulations of systems with multimodal  probability distributions are notoriously difficult because naive approaches tend to get trapped in one of the local minima for a significant amount of simulation time. Such a statement is true for statistical systems exhibiting first-order phase transitions such as the $Q\ge5$ two-dimensional Potts model \cite{potts} but also for more phenomenologically relevant models such as Lattice Quantum Chromodynamics \cite{Gattringer:2010zz}. In the latter case, the difficulty manifests itself as the well-known problem of freezing of topological charge in a single sector during the entire simulation. The problem is known since the beginning of computational sciences and many ideas have been put forward to circumvent that obstacle, in particular methods exploiting a number of concurrent simulations performed in different regions of the phase space equipped with mechanisms assuring that information can be shared and exchanged between different simulations \cite{PhysRevLett.57.2607,parallel_tempering}. More recent original proposals include Ref.~\cite{Albandea:2021lvl,Albandea:2021kwe} and Ref.~\cite{DelDebbio:2021qwf,DelDebbio:2021mts,Albandea:2022fky,Albandea:2023wgd} where it was suggested to employ neural networks as a realization of the trivializing map construction of Ref.~\cite{Luscher:2009eq}.

In this work, we investigate a new approach based on the incorporation of machine learning techniques into the simulation algorithms. It was recently proposed that  neural networks \cite{2019PhRvL.122h0602W,PhysRevD.100.034515,2020PhRvE.101b3304N,boltmann_generators_science,phiala,Albergo:2021bna,Albergo:2022qfi,Abbott:2022zhs,Abbott:2022hkm} can be used to enhance Monte-Carlo simulations of physical systems. In particular, Ref.~\cite{Hackett:2021idh} discussed the problem of multi-modal probability distributions in this context. When properly trained, such devices can generate independent consecutive configurations close to the target Boltzmann's distribution which can be  used to build Markov chains. An important property of this approach is that it allows to estimate the free energy and other thermodynamic observables with a remarkable precision \cite{PhysRevLett.126.032001}. One of the goals of this work is to demonstrate this for the Potts model.

Simulations of systems with discrete degrees of freedom like Ising or Potts models are enhanced by the so-called {\em autoregressive} neural networks \cite{2019PhRvL.122h0602W} (a variation of dense neural network with a lower triangular weight matrix). Recently this architecture was used for systems with continuous degrees of freedom \cite{Wang_2022}. They have the unfortunate drawback of requiring $L^2$ invocations of the neural network for generating a single, two-dimensional $L\times L$, configuration. This leads to an overall $L^6$ scaling of the costs of generating configurations. 
In Ref. \cite{Bialas:2022qbs} this scaling was replaced with a milder $L^3$, which was made possible by replacing a single neural network with $L^2$ inputs of the original approach with a hierarchical autoregressive network (HAN) consisting of smaller networks. 

It was also suggested in Ref.~\cite{Bialas:2022qbs} that the training of smaller neural networks can be performed on systems with smaller sizes and then used as the initial condition for the training of larger systems further reducing the cost of training. In some sense, this can be viewed as an analog of the ensemble of simulations performed in the parallel tempering algorithm. In the latter, one simulates different temperatures because the barriers between different modes of the distributions tend to be smaller at high temperatures. In the former, one tries to rely on the fact that these very same barriers tend to be smaller in smaller finite volumes. It is the aim of this study to provide numerical evidence that such an approach works in practice.

In the following, we present the results of the application of the hierarchical autoregressive neural network algorithm to the two-dimensional Potts model with $Q=12$ states. This number of states was chosen because in this situation the model exhibits a first-order phase transition and one can easily show the bimodal nature of the probability distribution. We will use this system to benchmark our approach and compare its performance against the Wolff cluster algorithm \cite{WOLFF1989379}. The rest of the paper is organized as follows. In Section \ref{sec. II} we briefly define the simulated systems. In Section \ref{sec. III} we remind the definitions and the construction of the hierarchical approach and discuss several improvements which were crucial in performing the simulations for the Potts model. In Section \ref{sec. IV} we very briefly describe the reference results obtained using the cluster algorithm. The main results of our numerical experiments are discussed in Section \ref{sec. V}, whereas in Section \ref{sec. VI} we provide values of the free energy and entropy across the phase transition. We gather our conclusions and provide a discussion of our findings in Section \ref{sec. VII}. Additional technical details can be found in four appendices.

\section{Definition of simulated system}
\label{sec. II}

In this work, we concentrate on the $Q$-state Potts model on a square lattice with periodic boundary conditions. A given association of spin states to lattice sites is called configuration $\mathbf{s}=\{s^1,s^2,\ldots ,s^{L^2}\}$, where each $s^i$ can occupy one of Q states.
Typically, the energy of the Potts model is given by the Hamiltonian 
\begin{equation}
    H(\mathbf{s}) = - \sum_{\langle i,j \rangle} \delta_{s^i \,, s^j},
\label{Potts_hamilt}
\end{equation}
where $\langle i,j \rangle$ denotes all nearest-neighbor pairs of sites on the lattice  and $\delta_{s^i \,, s^j}=\{ 1 \ \textrm{when} \ s^i = s^j, 0 \ \textrm{otherwise}\} $  is the Kronecker delta.
The corresponding Boltzmann probability distribution is given by
\begin{align}
    p(\mathbf{s}) = \frac{1}{Z(\beta)} \exp(-\beta H(\mathbf{s})) \,,
    \label{eq_boltz_distr}
\end{align}
with the partition function $Z(\beta)=\sum_{\mathbf{s}} \exp(-\beta H(\mathbf{s}))$, whose value is usually not known. $\beta$ is the dimensionless inverse temperature. It has been shown \cite{Baxter_1973} that for $Q \le 4$ the model has a second-order phase transition, whereas for $Q>4$ it exhibits a first-order phase transition. The infinite volume critical inverse temperature is known to be \cite{Baxter_1973}
\begin{equation}
    \beta_c(Q) = \ln(1 + \sqrt{Q}).
\end{equation}
In the following, we often use the rescaled parameter $\beta/\beta_c(Q)$ to tell how far from the phase transition the simulations were performed. However, note that in a finite volume the corresponding crossover transition happens at $\beta < \beta_c$. The order parameter which can be used to probe the phase of the system is the mean magnetization,
\begin{equation}
    m(\mathbf{s},Q) = \frac{1}{V} \max_{1\le q\le Q} \sum_{i=1}^{V} \delta_{q \,, s^i},
\end{equation}
where $V=L^2$ is the volume of our system.

By construction $\frac{1}{Q} \le m(\mathbf{s},Q) \le 1$. We will be often interested in statistical averages such as
\begin{equation}
    \avg{ m } \equiv \sum_{\mathbf{s}} p(\mathbf{s}) m(\mathbf{s}),
\end{equation}
where we have omitted the $Q$ dependence. We will estimate $\avg{ m }$ from a finite set of configurations,
\begin{equation}
    \avg{ m } \approx \frac{1}{N} \sum_{i=1}^N  m(\mathbf{s}_i), \qquad \mathbf{s}_i \sim p(\mathbf{s}),
\end{equation}
where by $\mathbf{s}_i \sim p(\mathbf{s})$ we mean that the set of $N$ configurations has been generated with the probability distribution $p(\mathbf{s})$. It is the aim of this work to provide a practical algorithm to estimate $\avg{ m }$ and other observables.

\section{Hierarchical Autoregressive Neural Networks}
\label{sec. III}

The main calculation will be performed with the algorithm proposed in Ref.~\cite{Bialas:2022qbs}. It employs an autoregressive neural network trained to approximate the target probability $p(\mathbf{s})$. As explained in the original proposal of Ref.~\cite{2019PhRvL.122h0602W}, the neural network is presented with a spin configuration, and as output, it returns the values of conditional probabilities for each spin. Hence, if the probability $p(\mathbf{s})$ is factorized as
\begin{equation}
    p(\mathbf{s}) = p(s^1) \prod_{i=2}^{L^2} p(s^i|s^1,s^2,\dots,s^{i-1}),
\end{equation}
then the $j$ output of the neural network corresponds to the approximation $q_{\theta}(s^j|s^0,\dots,s^{j-1})$ of $p(\mathbf{s}^j|\mathbf{s}^0,\dots,\mathbf{s}^{j-1})$ in such a way that the product of all outputs reproduces the configuration probability $q_{\theta}(\mathbf{s})$,
\begin{equation}
    q_{\theta}(\mathbf{s}) = q_{\theta}(s^1) \prod_{i=2}^{L^2} q_{\theta}(s^i|s^1,s^2,\dots,s^{i-1}) \approx p(\mathbf{s}).
    \label{eq. factorization}
\end{equation}
It was proposed in Ref.~\cite{2019PhRvL.122h0602W} to employ a single neural network of width $L^2$ to calculate all factors in Eq.~\eqref{eq. factorization}. Such an approach has, however, a prohibitive scaling with the system size $L$ not only from the point of view of the numerical cost of generating one configuration but also from the decreasing efficiency of training for larger neural networks. The latter is understood as the number of configurations needed to be generated to train the network to given quality (measured for example by the effective sample size). As a remedy to this problem, Ref.~\cite{Bialas:2022qbs}  discussed a hierarchical approach where the single large neural network was replaced by a hierarchy of smaller neural networks. This was possible due to the fact that the nearest-neighbor interactions in the studied model  Eq.~\eqref{Potts_hamilt} induce a restrictive pattern of dependencies of the conditional probabilities, i.e. the probability of a given spin depends conditionally on the values of spins on a closed contour enclosing that spin only; without any dependence on the spins outside of that contour (result known also as the Hammersley-Clifford theorem in the literature \cite{Hammersley-Clifford, Clifford90markovrandom}). 

\begin{figure*}
\begin{center}
\includegraphics[width=0.25\textwidth, angle=0]{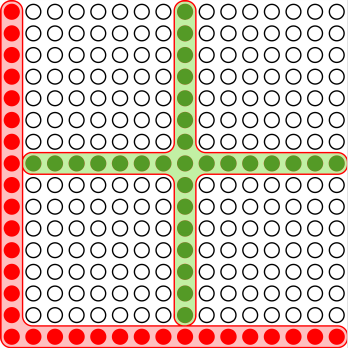}
\hspace{1cm}
\includegraphics[width=0.25\textwidth, angle=0]{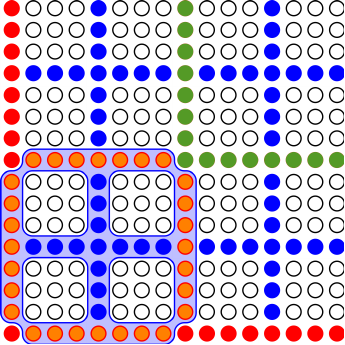}
\end{center}
\caption{Sketch of the partitioning of a $L \times L=16 \times 16$ lattice. Left panel: border spins are shown in red and the first set of interior spins in green. The probabilities of the green spins depend conditionally on all red spins. In the previous proposal of Ref.~\cite{Bialas:2022qbs} red and green spins were generated by a single neural network. Right panel: subsequent stages of spin generation proceed in the same way as described in Ref.~\cite{Bialas:2022qbs}. \label{fig:sketch}}
\end{figure*}

In practical terms, we recursively divide the spins on the square lattice with periodic boundary conditions into two classes: boundary and interior spins. The first iteration of such partitioning is demonstrated in the left panel of Fig.~\ref{fig:sketch}: within the  original lattice, we define the first boundary spins marked in red and the interior spins in a shape of a green cross. We chose this shape so that in each of the newly formed squares with fixed green or red boundaries we can repeat the same division. The second step is shown in the right panel of that Figure. We subsequently define the second set of interior spins marked in blue, which are again shaped as a cross. This procedure can be iterated until only one spin is left inside a square. To each class of spins, a separate autoregressive neural network is associated, i.e. all sets of blue spins in Fig.~\ref{fig:sketch} are generated iteratively by one neural network. The one used to generate the first set of red boundary spins has a traditional architecture as discussed in Ref.~\cite{Bialas:2022qbs}. The neural networks on all remaining sets of spins must account for the conditional dependency on the values of the boundary spins, and hence have a rectangular weight matrix: the number of input neurons is given by the sum of the boundary and interior spins, while the output is given by the number of interior spins they are supposed to generate. We show such an architecture explicitly in Fig.~\ref{fig:sketch architecture}.

As was shown in Ref.~\cite{Bialas:2022qbs}, the cost of generating a single configuration scaled as $L^6$ in the original approach is reduced to $L^3$. Moreover, the hierarchical construction also significantly improved the training efficiency. In the present work, we build upon this improvement and show that it allows studying larger systems with more degrees of freedom with excellent quality. We demonstrate this using the Potts model with $Q=12$ on a $L=32$ square lattice with periodic boundary conditions as a test bed. We simulated the model in the close vicinity of the first-order phase transition where many algorithms suffer from ergodicity difficulties. Below we describe several further improvements which contributed to the success of the approach and provide details of our training strategy.

\subsection{Generalization to the $Q$ state Potts model}

\begin{figure}
\begin{center}
\includegraphics[width=0.45\textwidth]{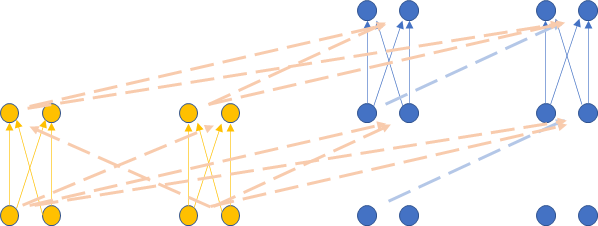}
\end{center}
\caption{Sketch of the autoregressive neural network architecture with conditional connections. For simplicity, we show the case of blocks of size $Q=2$. It is a generalization of the architecture of conditional autoregressive neural network described in Ref.~\cite{Bialas:2022qbs}. The blue part is an autoregressive neural network, and the yellow part implements the conditional dependency of the boundary spins. Solid lines show all-to-all connections between neurons belonging to two blocks. Dashed lines are drawn between blocks and summarize all-to-all connections between participating neurons. \label{fig:sketch architecture}}
\end{figure}

In order to accommodate additional states on each lattice site we have to modify the architecture of the neural networks present in the hierarchical approach. The main modification is the use of one-hot encoding in order to control the probabilities of each state on each site. Hence, each neuron in the neural networks from Ref.~\cite{Bialas:2022qbs} is promoted to a block of $Q$ neurons on the input and output layers. In order to have output probabilities properly normalized, we introduce a block softmax layer as the last layer. We preserve the autoregressive structure of connections between individual blocks since it reflects the possible dependencies of conditional probabilities on states at different lattice sites. We allow for all-to-all connectivity between neurons inside a block-to-block connection. We illustrate that in the sketch presented in Fig.~\ref{fig:sketch architecture}. The yellow neurons correspond to the conditional dependency on the boundary spins, while the blue neurons control the spins whose probabilities we are modeling. In the presented sketch we show the situation for $Q=2$, i.e. each block is made out of two neurons. We tested our implementation for this particular case of $Q=2$ since it corresponds to the Ising model which is analytically solvable. We provide an example of such tests in Appendix \ref{ap. potts}.

\subsection{Pretraining}

One of the main advantages of the hierarchical approach is that all, except the largest, neural networks can be used in simulations of systems with larger extents. We call the method of using neural networks on the lower levels of the hierarchy trained using smaller system sizes as \emph{pretraining}. 

\begin{figure*}
\begin{center}
  \includegraphics[width=0.495\textwidth]{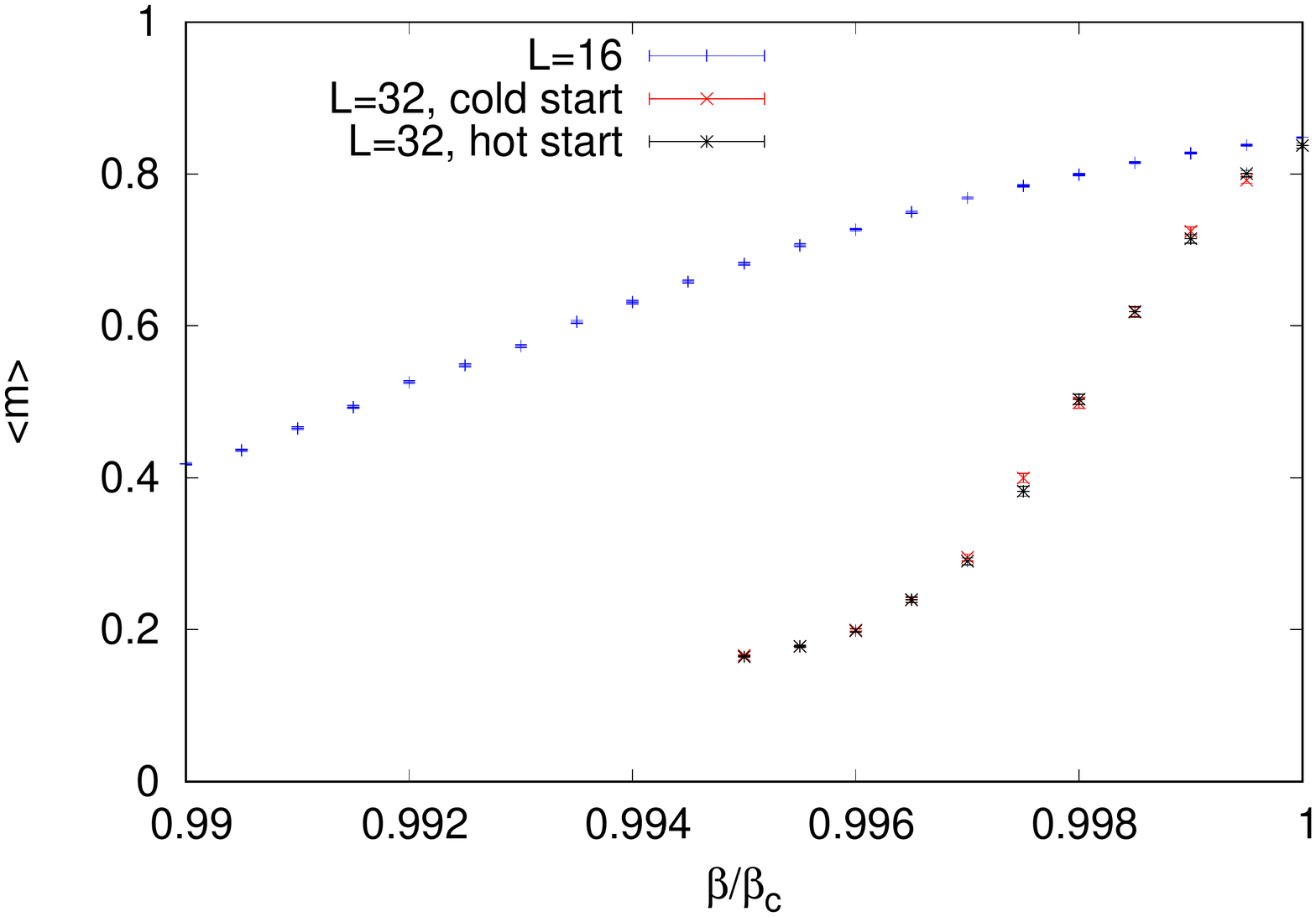} 
  \includegraphics[width=0.495\textwidth]{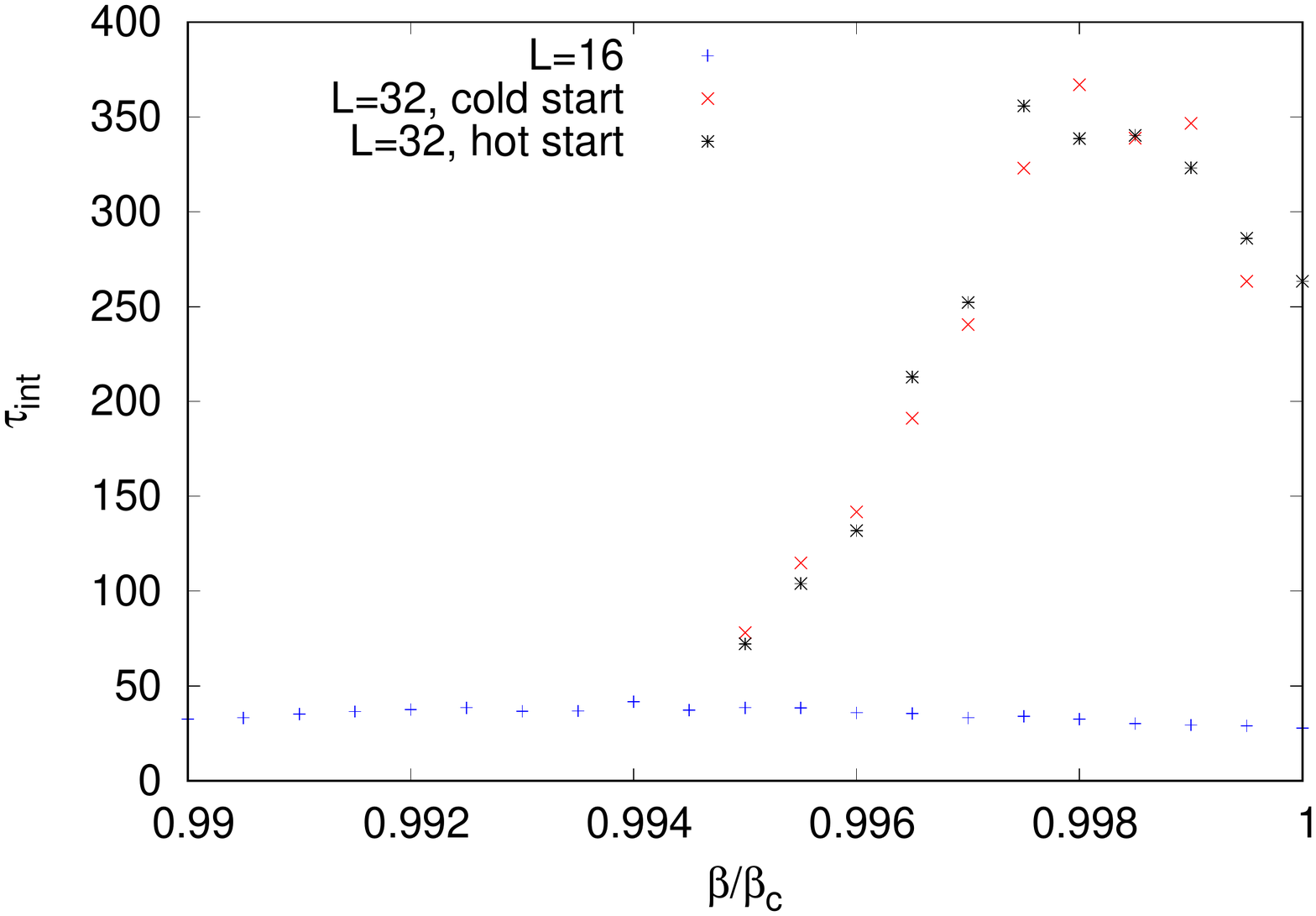} 
\end{center}
\caption{Results obtained using the cluster algorithm. The left panel shows the dependence of the magnetization for $L=16$ and $L=32$ across the phase transition, whereas the right panel shows the size of integrated autocorrelation time, $\tau_{int}$. "cold" and "hot" labels refer to two independent Markov chains started from a cold or a hot configuration, respectively. Compatible results testify that both chains have been properly thermalized and the amount of tunneling between the two energy minima is comparable in both of them. \label{fig: wolff}}
\end{figure*}

At first sight, trained neural networks at the lower levels of the hierarchy could be used in larger systems without further training. This is because networks like the one depicted in the right panel of Fig.~\ref{fig:sketch architecture} are completely independent of the rest of the system (all the dependence is captured by the conditional dependence on the boundary spins). However, this is true only for perfectly trained networks. In practice, the network is trained on some distribution of possible boundary configurations (red spins). If we use that neural network in a simulation of a larger system, the boundary spins may be distributed according to some other distribution and the final results may turn out to be poor. Such is the case when we are close to the phase transition. As correlations grow, they induce significant finite-volume effects. An example of such finite volume effect is anticipated in Fig.~\ref{fig: wolff} where we show the magnetization for the Potts model around the phase transition for $L=16$ and $L=32$. For a fixed value of $\beta<1$ we see a large difference in magnetization between the systems of these two extents. The neural networks trained at $\beta/\beta_c=0.995$ and $L=16$ will encounter configurations that are mostly ordered as the mean magnetization is around $0.7$. The same neural network used in the simulation at $L=32$ should better approximate conditional probabilities for unordered boundaries as the magnetization for that system size at the same inverse temperature is only equal to $0.2$. It is then not obvious how to transfer neural networks between different system sizes in this region close to the phase transition. We expect that such finite volume effects decrease as $L^{-\alpha}$, $\alpha \ge 1$, hence this problem should become less important as we reach larger system sizes. In practice we take networks trained at $\beta$ for which $\langle m \rangle \approx 0.5$, insert them into larger systems and train them further.

\subsection{Imposition of global symmetries}

We observe that the quality of neural network training is better if we impose global symmetries during training \cite{2019PhRvL.122h0602W,Bialas:2022qbs, Bialas:2021bei}.  This can be achieved by defining a symmetrized probability: for each generated configuration its probability is replaced by the  probability averaged over the configuration's symmetry images,
\begin{equation}
    q_{\theta}(\mathbf{s}) \rightarrow \bar{q}_{\theta}(\mathbf{s}) = \frac{1}{M}\sum_{i=1}^M q_{\theta}( h_i(\mathbf{s})),
\end{equation}
where $h_i$, $i=1,\dots, M$ are symmetry operators defined as transformations of the configuration space which keep the energy unchanged.

The significance of this step comes from the fact that the neural network will \emph{a priori} give different probabilities to equivalent configurations,
\begin{equation}
    q_{\theta}(\mathbf{s}) \ne q_{\theta}(h_i(\mathbf{s})).
\end{equation}
By using the averaged probability $\bar{q}_{\theta}$ we force the resulting probabilities to be symmetric.

In our work, we use the Kullback--Leibler (KL) divergence as the loss function. Including the symmetrization, its definition reads
\begin{align}
    \bar{D}_\textrm{KL} (q_\theta | p) &= \sum_{k=1}^N \bar{q}_\theta(\mathbf{s}_k) \, \ln \left(\frac{\bar{q}_\theta(\mathbf{s}_k)}{p(\mathbf{s}_k)}\right).
    \label{eq:KL_loss_sym} 
\end{align}
 One has to keep in mind that the configurations are generated by the network from the distribution $q_\theta(\b s)$. This distribution does not respect symmetries and provides a worse approximation to $p(\b s)$ than $\bar q_\theta(\b s)$. However, as long as we are using observables that are invariant under the symmetries $h_i$, we are allowed to use $\bar q_{\theta}$ in all our calculations (see Eqs.~(\ref{eq. w bar}) and (\ref{accept_rej_condition})). This is explained in detail in Appendix~\ref{app:symm}. 

In this work, we consider a varying set of symmetry operators which include translations in both directions, reflection $x \leftrightarrow y$ mirror symmetry as well as permutations in the internal space of $Q$ states. The latter corresponds to the $\mathbf{s} \rightarrow -\mathbf{s}$ spin flip symmetry of the Ising model. We increase the number $M$ of included symmetries with the increasing number of training epochs.

\section{Benchmark simulations with the Wolff cluster algorithm}
\label{sec. IV}

We compare our results with the Wolff cluster algorithm \cite{WOLFF1989379} which is considered as a reference algorithm for Ising and Potts models. 
The basic step of that algorithm is a single cluster flip: i.e. for a site chosen at random a single cluster is constructed around it. All the spins in this cluster are changed to a new, randomly chosen state (different from the current state of the cluster’s spins). We perform $L^2$ such flips in one sweep. We measure all observables every sweep and the resulting autocorrelation time is in units of sweeps.

In Fig.~\ref{fig: wolff} we present results obtained using the Wolff cluster algorithm \cite{WOLFF1989379}. The left panel shows the magnetization per spin as a function of the inverse temperature for two system sizes: $L=16$ and $L=32$. Each simulation consisted of $10^6$ sweeps and the additional first $10^5$ sweeps were discarded. The phase transition becomes sharper and sharper with increasing volume. For the larger size, we show results from two runs, one starting from a cold configuration where all spins were initialized in the same state and a hot configuration where initial spins we set at random. Both results agree within their statistical precision which includes a factor proportional to the square root of the integrated autocorrelation time. In the right panel, we show the integrated autocorrelation time itself for the smaller and larger system evaluated by integrating the positive part of the autocorrelation function. We see that it drastically increases at the phase transition practically prohibiting any simulation at larger system sizes, i.e. $L \ge 64$.

\begin{figure*}
\begin{center}
  \includegraphics[width=0.495\textwidth]{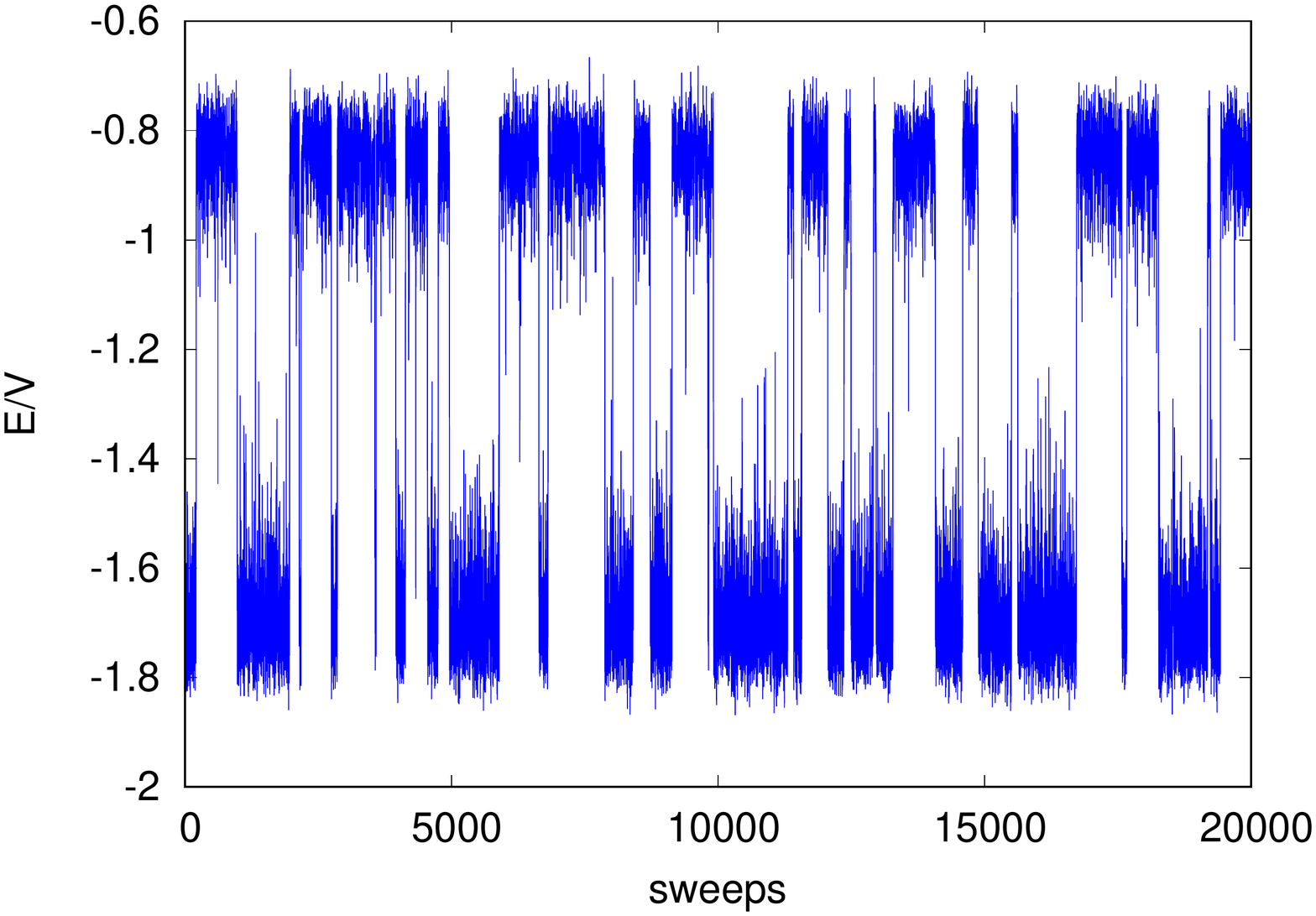} 
  \includegraphics[width=0.495\textwidth]{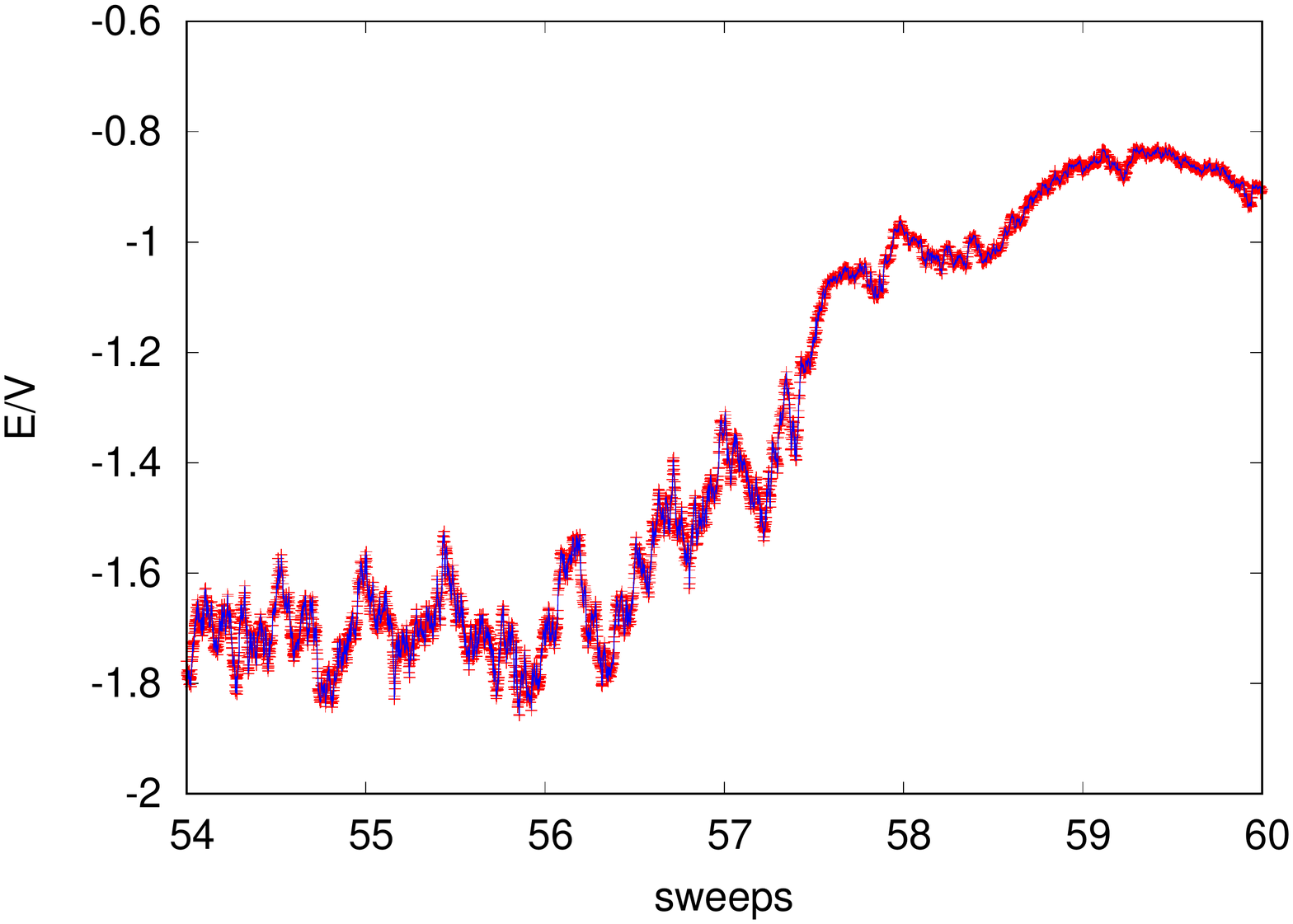} 
\end{center}
\caption{Example of a Markov chain at $\beta/\beta_c = 0.998$ for $L=32$. In the left panel we show the energy recorded every sweep ($L^2$ cluster flips), whereas in the panel on the right we show the energy recorded after every single cluster flip. The potential has two minima, one with configurations with energy around $-0.9$ and the second with configurations with energy around $-1.7$. The trajectory exhibits long periods of configurations that stay in one of the modes and cannot tunnel to the other mode. During the tunneling configurations with intermediate energies are visited along the path. This will not be the case in the HAN approach which we comment on more in the main text. \label{fig: trajectory}}
\end{figure*}

In Fig.~\ref{fig: trajectory} we show  a small piece of history of one of the Markov chains generated using the cluster algorithm at $\beta/\beta_c=0.998$ for $L=32$. We use it to demonstrate the origin of large autocorrelation times. One clearly sees that for long periods the simulation is confined in one of the modes before it travels to the other one where again it remains for a considerable amount of time. The transition between the two modes is captured on the right panel of that figure and we would like to stress that the transition does not have a nature of a sudden jump but is composed of a series of small steps hence covering also configurations that do not belong to either of the modes. We will come back to this issue when we discuss similar results for the neural Markov Chain Monte Carlo algorithm in the next section.

\section{Simulations across first order phase transition for $Q=12$}
\label{sec. V}

This section contains the main results of our work. Using our hierarchical approach we managed to train the corresponding sets of neural networks for a relatively large system size of $L=32$ for the Potts model with $Q=12$ in the vicinity of the phase transition. Based on the data from the left panel of Fig.~\ref{fig: wolff} we have chosen 7 temperatures where the mean magnetization changes significantly: $\beta/\beta_c=0.9965,0.997,0.9975,0.998,0.9985,0.999$ and $0.9995$ for which our neural network models were trained independently. Some details of the training strategy are provided in Appendix \ref{ap. quality}.

\begin{table}[]
    \centering
    \begin{tabular}{|c|c|}
    \hline
    $\beta/\beta_c$ & ESS \\
    \hline 
    0.9965 & 0.28 \\
    0.9970 & 0.53 \\
    0.9975 & 0.54 \\
    0.9985 & 0.49 \\
    0.9980 & 0.51 \\
    0.9990 & 0.68 \\   
    0.9995 & 0.70 \\   
    \hline
    \end{tabular}
    \caption{The values of ESS for the neural networks trained at seven inverse temperatures used in this paper.}
    \label{tab: ess values}
\end{table}

Following Ref.~\cite{Liu,phiala} we use the effective sample size (ESS) observable to quantify the quality of the training,
\begin{equation}
    \textrm{ESS}= \frac{\avg{\hat{\bar{w}}}_{q_\theta}^2}{\avg{\hat{\bar{w}}^2}_{q_\theta}},
\label{ESS_definition}
\end{equation}
where 
\begin{equation}
    \hat{\bar{w}}(\mathbf{s}) = \frac{e^{-\beta H(\b s)}}{\bar{q}_{\theta}(\mathbf{s})}.
    \label{eq. w bar}
\end{equation}
are the (unnormalized) importance weights.
This quantity is normalized such that $\textrm{ESS} \in [0,1]$ and for perfectly matched distributions one has $\textrm{ESS}=1$. For each of the seven temperatures, we generated $10^6$ configurations once the training has reached acceptable quality. We estimated the ESS and we gathered them in Tab.~\ref{tab: ess values}.  It is remarkable that most of the  values of ESS are greater than $0.5$ while the values at the beginning of the training were less than $10^{-4}$.

\begin{figure*}
\begin{center}
  \includegraphics[width=0.495\textwidth]{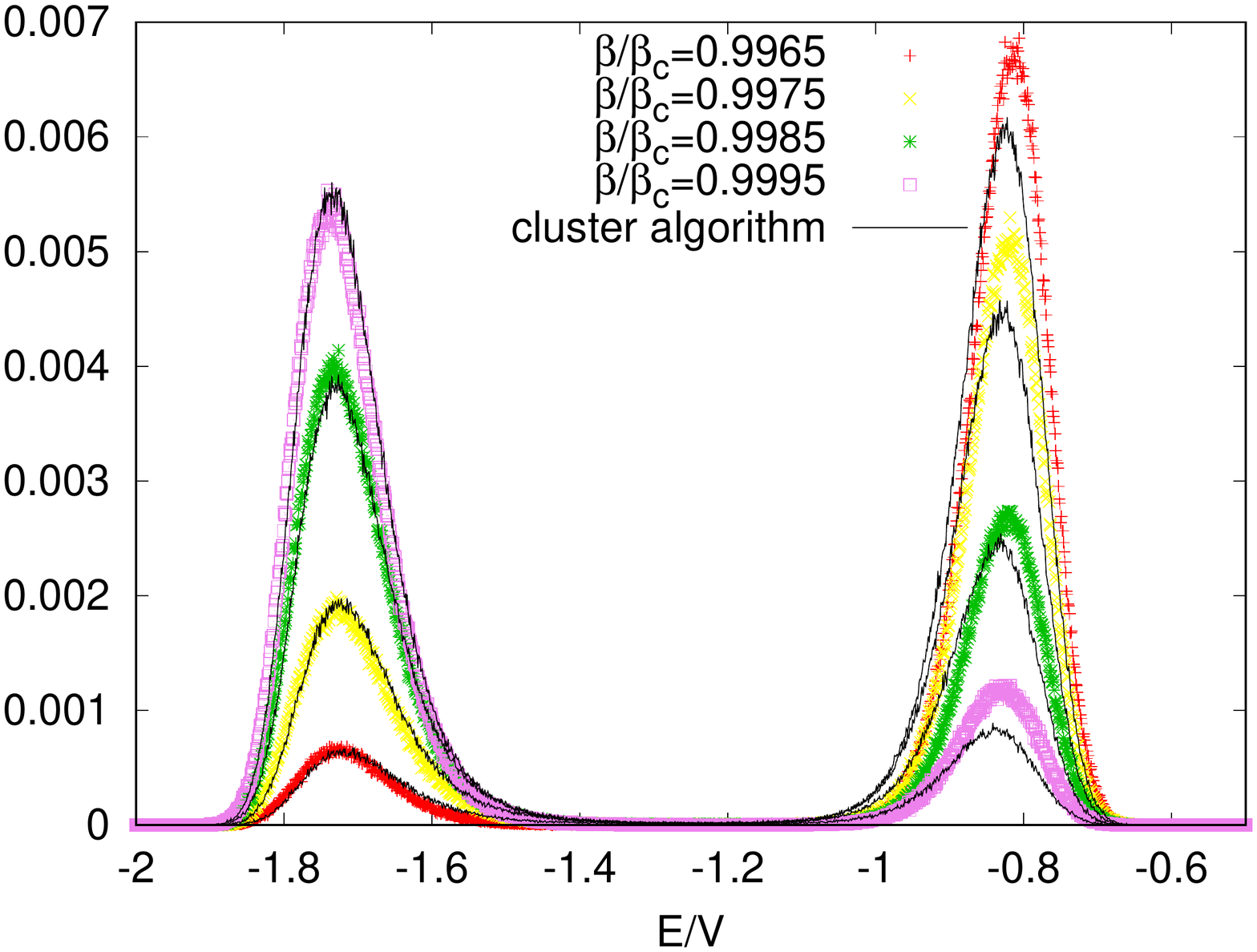} 
  \includegraphics[width=0.495\textwidth]{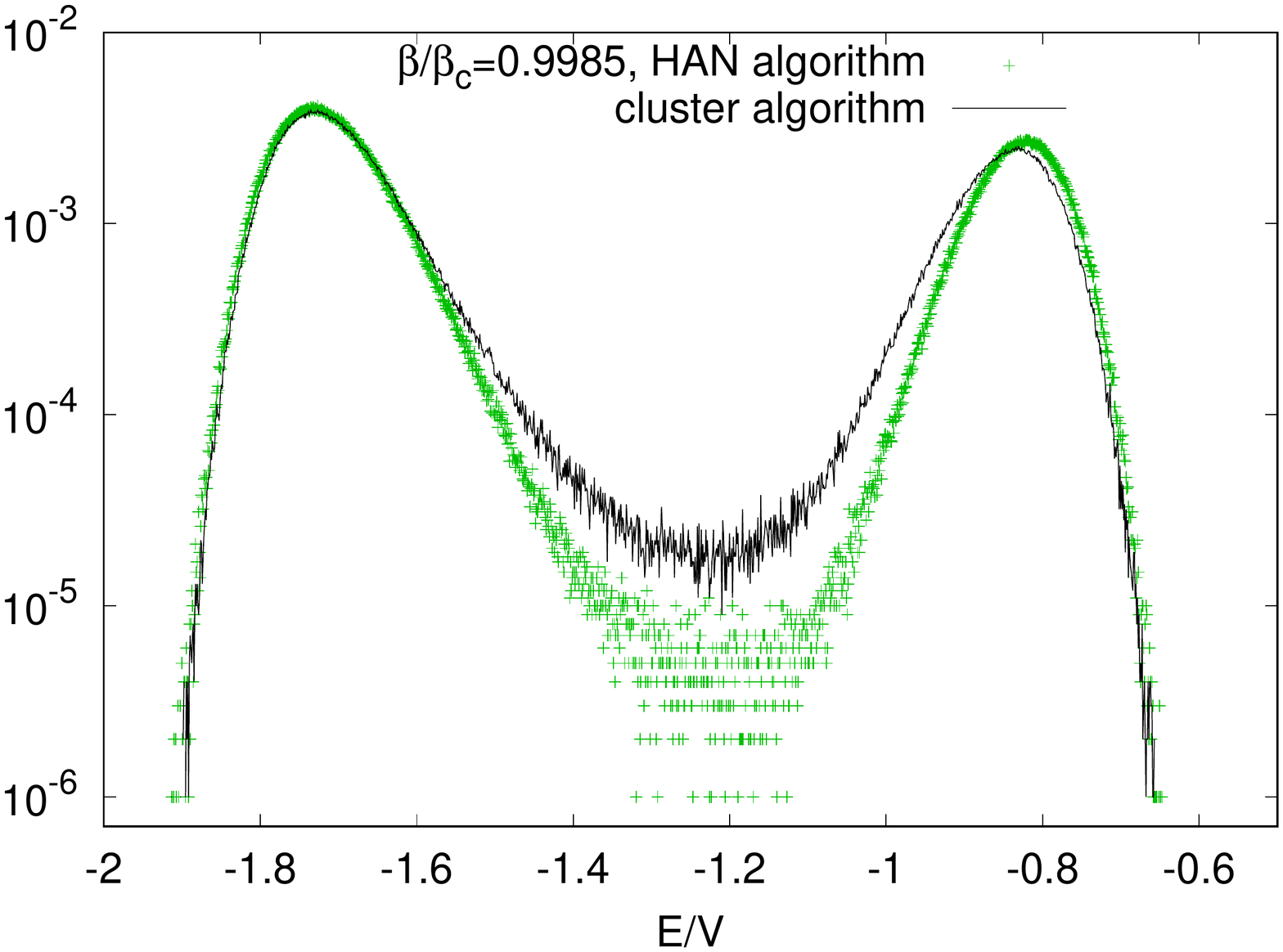} 
  \end{center}
\caption{Comparison of energy density histograms obtained directly from the HAN approach and from the cluster algorithm.\label{fig:histogram_energy}}
\end{figure*}

We use the previously generated $10^6$ configurations to produce histograms of the energy density, i.e. we count how many configurations have been generated with a given energy. This can be compared to similar results obtained from a corresponding number of $10^6$ configurations generated using the Wolff cluster algorithm. We show such a comparison in the left panel of Fig.~\ref{fig:histogram_energy} for four inverse temperatures to make the plot more readable. Cluster algorithm results are shown with black solid lines, whereas color data come from the HAN approach. One clearly sees that the system has two modes, one at $E/V \approx -1.7$ and the second at $E/V \approx -0.9$. The interesting observation is that both methods have sampled both of the modes and that the contribution of each mode changes as the temperature is changed. It is also clearly visible that the neural networks are sampling a bit too often configurations in the right mode. It may be worth looking closer at the results in between the two modes, i.e. at energies $E/V \approx -1.2$. Since the counts in this region are relatively small, we show the histogram for $\beta/\beta_c=0.9985$ on a logarithmic scale in the right panel of that figure. It is evident that the neural network undersamples in this transition region.

\begin{figure*}
\begin{center}
   \includegraphics[width=0.495\textwidth]{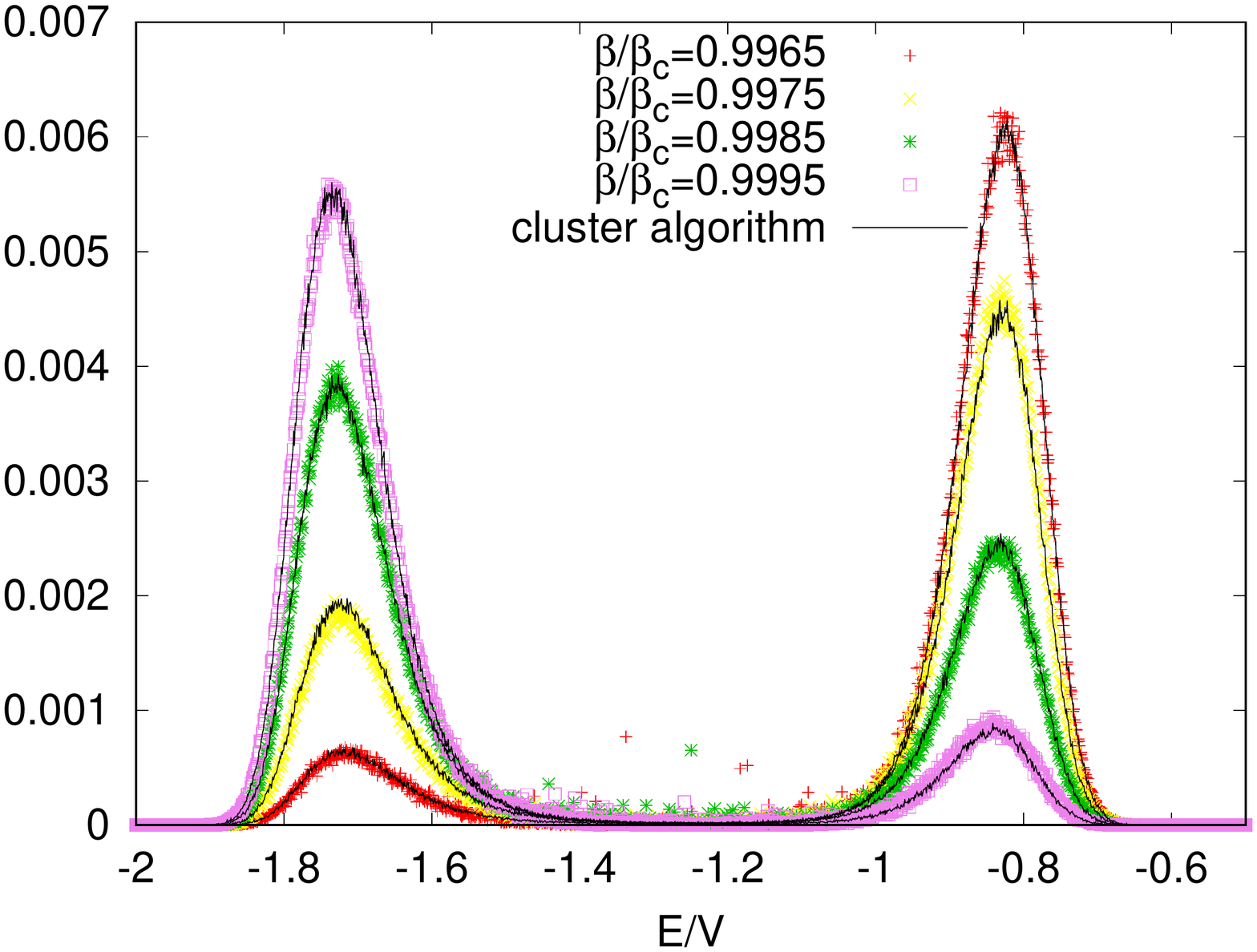} 
    \includegraphics[width=0.495\textwidth]{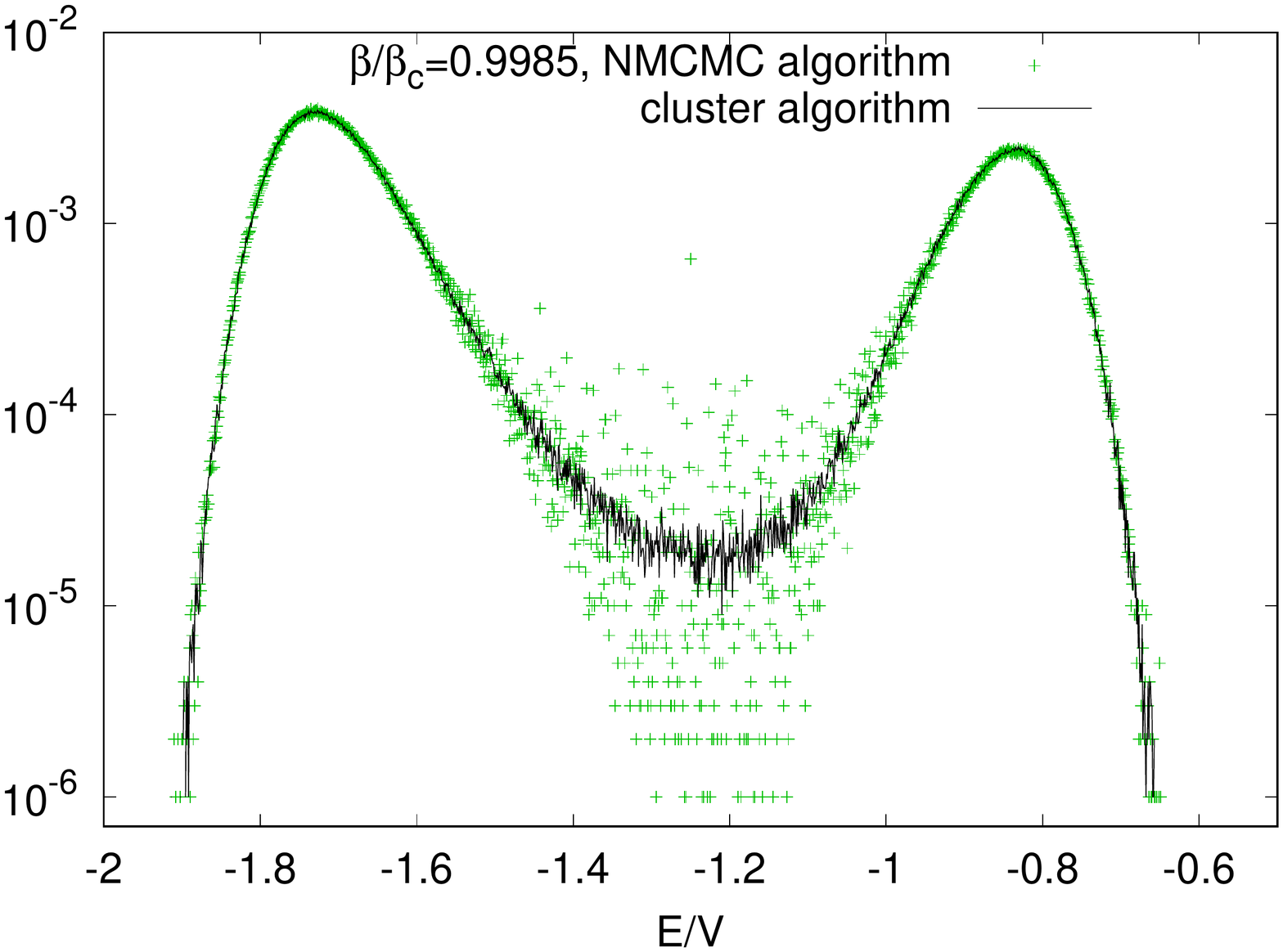} 
\end{center}
\caption{Comparison of energy density histograms obtained from the HAN NMCMC approach and from the cluster algorithm.\label{fig:histogram_energy_nmcmc}}
\end{figure*}

In the literature, there exist two approaches suitable to account for the difference between $\bar{q}_{\theta}$ and the target probability $p$ \cite{PhysRevD.100.034515,2020PhRvE.101b3304N}. In the first one, named Neural Importance Sampling (NIS), the mean value of an observable $\mathcal{O}$ is given by:
\begin{equation}
    \avg{ \mathcal{O} } = \frac{1}{N\hat{Z}} \sum_{i=1}^N  \hat{\bar{w}}(\mathbf{s}_i) \mathcal{O}(\mathbf{s}_i), \qquad \mathbf{s}_i \sim q_{\theta},
\end{equation}
where the importance ratios are defined in Eq.~(\ref{eq. w bar}) and the estimate of the partition function is given by:
\begin{equation}
    \hat{Z} = \frac{1}{N}\sum_{i=1}^N \hat{\bar{w}}(\mathbf{s}_i), \qquad \mathbf{s}_i \sim q_{\theta}.
    \label{Zhat def}
\end{equation}

In the Neural Markov Chain Monte Carlo (NMCMC) approach, one applies the accept-reject step to configurations generated by the network. The acceptance probability for the configuration $\mathbf{s}_{k+1}$ is given by
\begin{equation}
    \min \left( 1, \frac{p(\mathbf{s}_{k+1}) \bar q_\theta(\mathbf{s}_{k})}{p(\mathbf{s}_{k}) \bar q_\theta(\mathbf{s}_{k+1}) }   \right) = 
    \min \left( 1, \frac{\hat{\bar{w}}(\mathbf{s}_{k+1}) }{\hat{\bar{w}}(\mathbf{s}_{k}) }   \right)
    ,
\label{accept_rej_condition}
\end{equation}
where $\mathbf{s}_{k}$ is a previous configuration in the Markov Chain.

As explained in Ref.~\cite{2020PhRvE.101b3304N} both NIS and NMCMC are exact and provide correct results within their statistical uncertainties. 

As a first step in the comparison of results we look at the energy density histograms discussed above, now after the accept/reject step (very similar conclusions can be drawn from weighted histograms in the NIS approach, see Appendix~\ref{app:energy-density}). We compare them to the cluster algorithm in Fig.~\ref{fig:histogram_energy_nmcmc}. In the left panel, we notice that the color data reproduces to a high precision the black line from the cluster algorithm. This is also clearly visible in the right panel on the logarithmic scale. 

However, looking at the region of $E/V\approx -1.2$ at the logarithmic scale, we notice that while NMCMC data correctly approximates the results from the cluster algorithm, it has a much bigger variance. This is due to the severe undersampling in this region. Looking back at  Figure ~\ref{fig:histogram_energy} we see that for some  energy values only one configuration was generated. While  the accept/reject mechanism or NIS  will correct this, it  produces a large error because of poor statistics.

\begin{figure}
\begin{center}
  \includegraphics[width=0.495\textwidth]{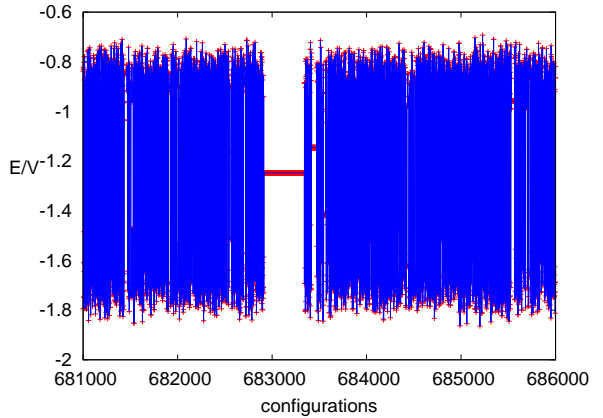} 
\end{center}
\caption{Example of a Markov chain at $\beta/\beta_c = 0.998$ for $L=32$ from HAN NMCMC. Underestimated probability of configurations with intermediate energies leads to long sequences of rejected trajectories. As a result, the probabilities of these configurations lying between the two minima have large uncertainties in a Markov chain of a fixed length. The figure is concentrated on the longest rejection sequence of 481 configurations in the $10^6$ Markov chain. \label{fig: trajectory nmcmc}}
\end{figure}

\begin{figure*}
\begin{center}
    \includegraphics[width=0.49\textwidth]{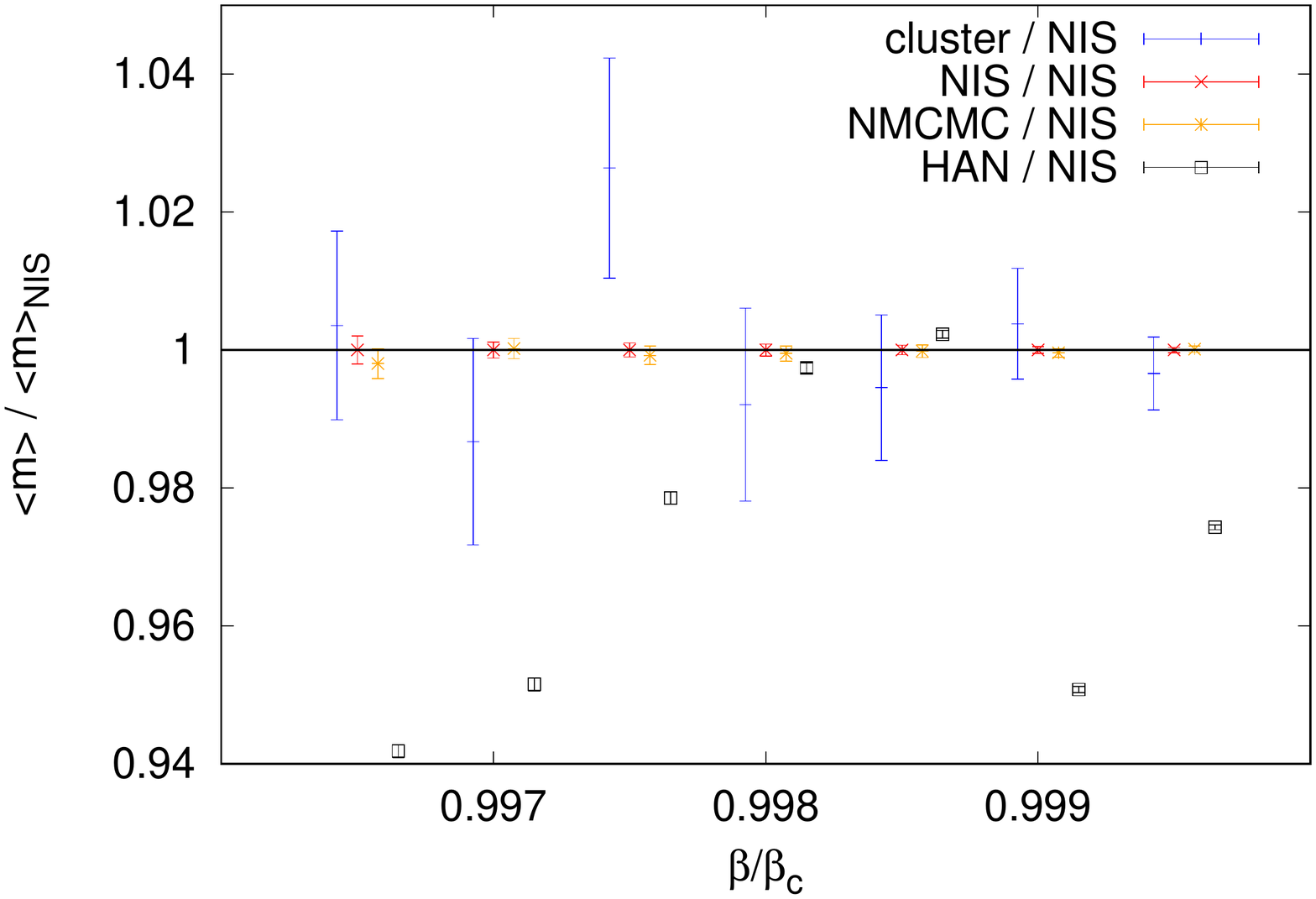} 
    \includegraphics[width=0.49\textwidth]{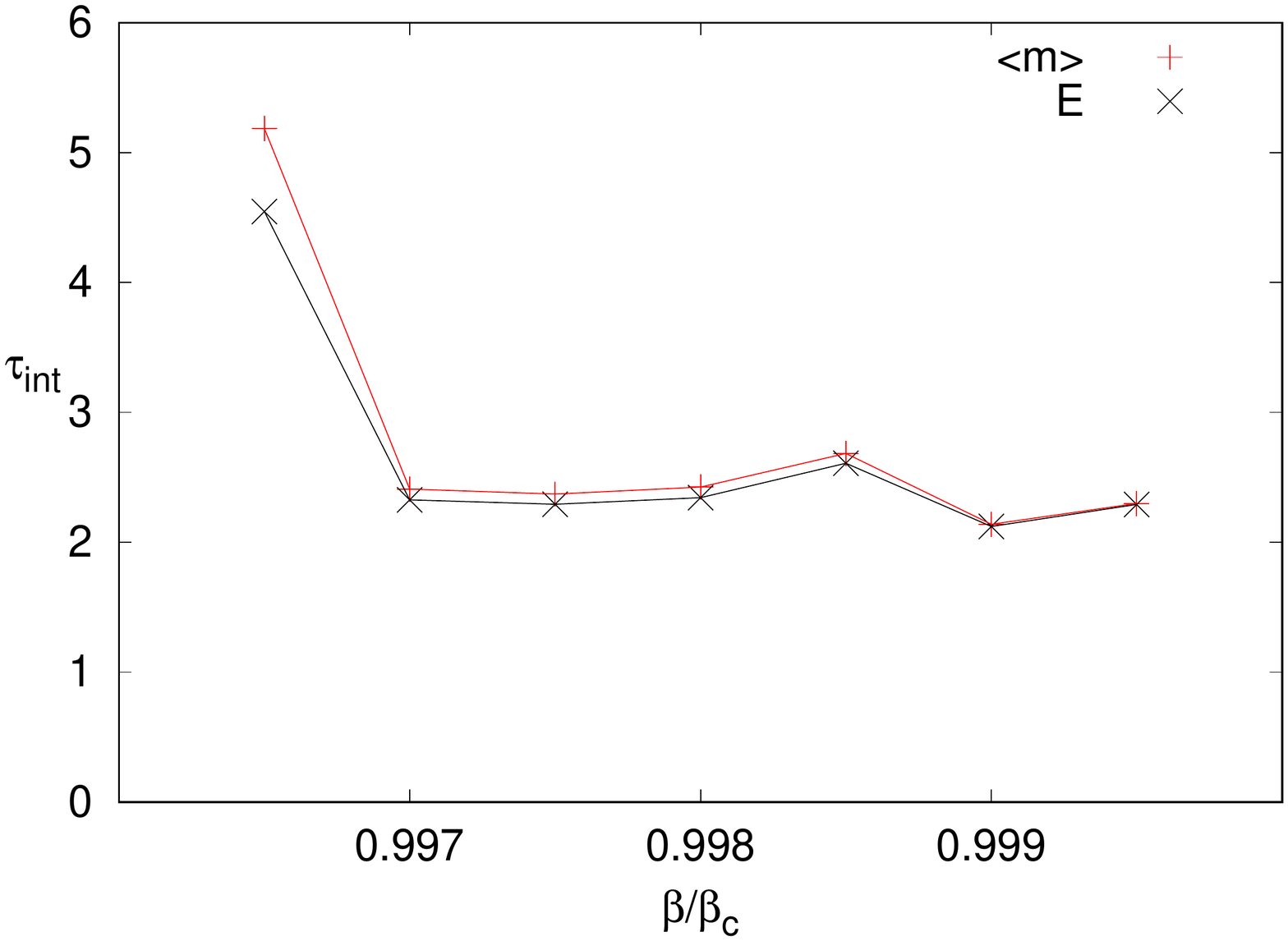}
\end{center}
\caption{Left panel: Comparison of mean magnetization obtained with the HAN, NIS, NMCMC, and the cluster algorithm. All mean values and errors were divided by the NIS mean value. Right panel: Integrated autocorrelation time of the NMCMC algorithm based on the $\langle m \rangle$ and $E$ data as a function of $\beta$ for $L=32$. No significant critical slowing down phenomenon can be seen in the vicinity of the phase transition.\label{fig:autocorrelation magnetization}}
\end{figure*}

Undersampling also generates a non-trivial, i.e. significantly larger than $1$, autocorrelation time of the NMCMC Markov chain. Configurations with $\q(\v s)\ll p(\v s)$ will have a small probability Eq.~\eqref{accept_rej_condition} of being replaced by another configuration and thus will be retained for a large number of steps compensating for the undersampling.  This is exactly what we observe in Fig.~\ref{fig: trajectory nmcmc} where we show a piece of the NCMCM Markov Chain measuring the energy density on each accepted configuration. The neural network very nicely samples both modes, jumping independently into each of them. However, once a configuration from the transition region occurs, since its probability is underestimated compared to the target probability, it is retained for a considerable amount of steps.  
On the other hand, configurations located around the two peaks of the histogram, are much more probable to be accepted in accept/reject step due to the small difference between $p$ and $q_\theta$. 
In principle, rejections of both types of configurations, i.e. from the two modes and from the intermediate region, contribute to the overall autocorrelation time. However, we have checked that in the case of energy and magnetization, the autocorrelation time is mostly generated by  rejections of configurations representing the two modes. To show this, we removed configurations with $-1>E/V>-1.5$ (around 30k) from the set of configurations proposed by the neural network (which were $10^6$), we applied the accept/reject step to the remaining ones, and measured the autocorrelation time. For $\beta/\beta_c=0.998$ we obtained $\tau_{\textrm{int}}=2.29$, compared to $\tau_{\textrm{int}}=2.34$ obtained without the removal of intermediate configurations.
The maximal autocorrelation time measured in our NMCMC chains 
did not exceed $5$ (see Fig.~\ref{fig:autocorrelation magnetization}).

Having made sure that the HAN approach correctly samples from the bi-modal energy distribution of configurations we can compare the average results for $\langle m \rangle$ obtained using different algorithms and their statistical uncertainties. We do that for mean magnetization in Table~\ref{tab:est-comp}. We visualize these results in
Fig.~\ref{fig:autocorrelation magnetization} where we 
show the ratios of the values of the mean magnetization obtained using the four methods divided by the most precise result from the NIS method. The statistical uncertainties have been rescaled by NIS mean value, so they reflect directly the precision of each of the approaches. In all cases, we have used $10^6$ configurations. Evident agreement between the NIS, NMCMC, and the cluster algorithms can be seen, with the NIS and NMCMC approaches being much more precise. As we have anticipated in the right panel of Fig.~\ref{fig: wolff}, the autocorrelation time of the cluster algorithm is around 400 sweeps, while in the case of NMCMC $\tau_{\textrm{int}}$ is almost everywhere $\approx 2$  (see right panel of Fig.~\ref{fig:autocorrelation magnetization}) explaining the factor of more than 10 in the size of statistical uncertainties. The standalone HAN algorithm underestimates magnetization for most temperatures by up to 6\%. This is yet another visualization of imperfect training of the neural network and overestimating right mode in Fig.~\ref{fig:histogram_energy}.

\begin{table}
\centering
 \begin{tabular}{|c||c||c|c||c|}
\hline
$\beta/\beta_c$ & HAN & NIS  & NMCMC & Cluster \\
 \hline
0.9965	&  0.22438(23) &    0.23824(49)	&	0.23777(52) & 0.2391(33)  \\
0.9970  &  0.28468(28) &	0.29919(35)	&	0.29925(45)	& 0.2952(45)  \\
0.9975  &  0.38112(33) &	0.38947(40)	&	0.38917(51)	& 0.3998(62)  \\
0.9980	&  0.50126(37) &    0.50257(43)	&   0.50231(56)	& 0.4986(70)	\\
0.9985	&  0.62277(35) &    0.62134(43)	&   0.62124(58)	& 0.6179(66)	\\
0.9990	&  0.68628(33) &    0.72182(35)	&   0.72153(45)	& 0.7246(58)	\\
0.9995  &  0.77408(28) &    0.79451(30) &   0.79460(38)	& 0.7918(42)  \\
\hline
\end{tabular}
\caption{Numerical values for the mean magnetization and its statistical uncertainty obtained with HAN, NIS, NMCMC, and cluster algorithms. Ratios of these entries are shown in Fig.~\ref{fig:autocorrelation magnetization}.}
\label{tab:est-comp}
\end{table}

\begin{figure*}
\begin{center}
  \includegraphics[width=0.495\textwidth]{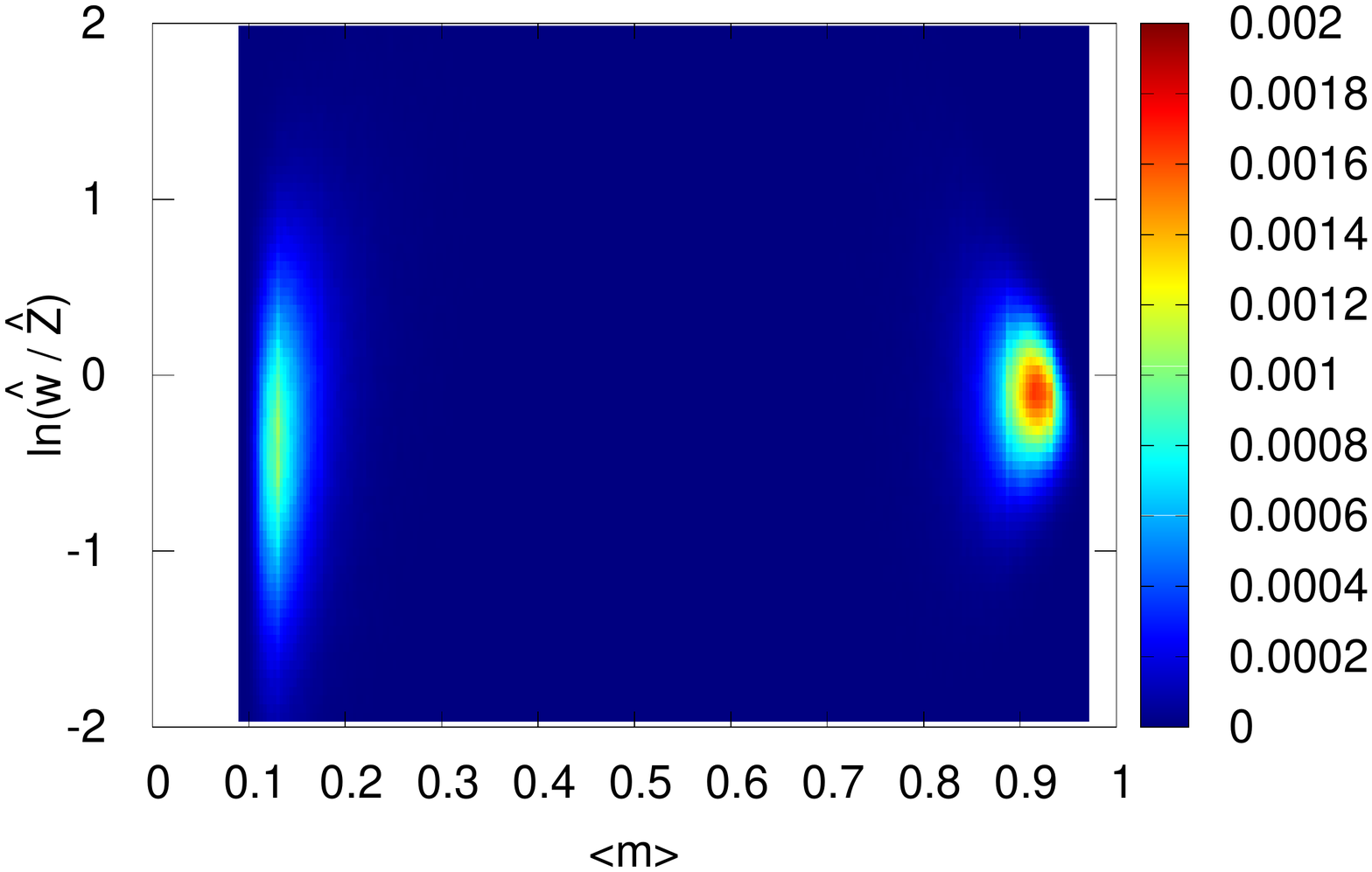} 
  \includegraphics[width=0.495\textwidth]{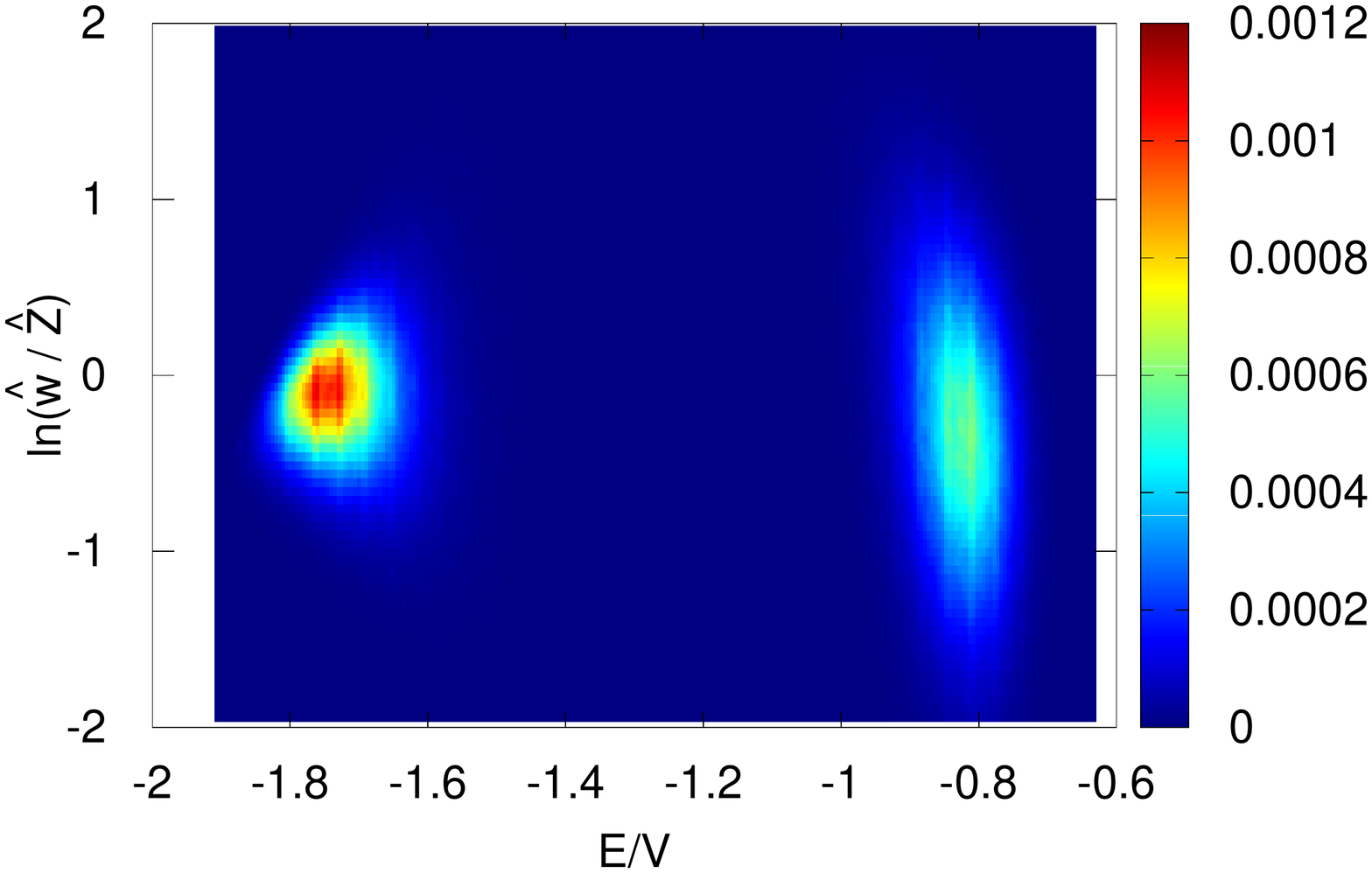} 
\end{center}
\caption{Normalized reweighting factors in the NIS approach against the magnetization (left panel) and energy (right panel) for $\beta/\beta_c = 0.998$ and $L=32$.\label{fig:histogram}}
\end{figure*}

Eventually, in order to check the stability of the NIS approach we investigate the distribution of the normalized reweighting factors of Eq.\eqref{eq. w bar}. We plot $\hat{\bar{w}}/\hat{Z}$ for all $10^6$ configurations generated at $\beta/\beta_c=0.998$ in Fig.~\ref{fig:histogram}. We depict $\ln(\hat{\bar{w}}/\hat{Z})$ against $\langle m \rangle$ and against $E/V$ on the left and right panels, respectively. We clearly identify two clusters corresponding to the two modes of energy and magnetization for this particular 
$\beta/\beta_c$. The majority of reweigthing factors have $\ln(\hat{\bar{w}}/\hat{Z}) \approx 0$ and are located in the mode $E/V \sim -1.8$. One can check that all remaining factors have relatively small values $|\ln(\hat{\bar{w}}/\hat{Z})| < 2$ which confirms that the reweighting procedure does not introduce significant statistical uncertainty.

We conclude that both the NIS and NMCMC approaches provide reliable results across the first-order phase transition and allow to obtain very precise results at comparable numerical costs.

\section{Precise thermodynamic state functions}
\label{sec. VI}

The possibility of generating configurations from the probability distribution $\q$ together with the values of the probability $\qb$ for each configuration allows estimating thermodynamical observables and the partition function $Z(\beta)$. Following Ref.~\cite{2020PhRvE.101b3304N} we use the NIS estimators for the partition function, free energy and entropy which include reweighting factors correcting the probability distribution $q_{\theta}$ back to the target $p$.
Reweighted free energy and entropy are given by
\begin{equation}
{F}_{\textrm{NIS}} = -\frac{1}{\beta} \ln \hat{Z} 
\end{equation}
and
\begin{equation}
{S}_{\textrm{NIS}} = \frac{1}{N \hat{Z}} \sum_{i=1}^N \hat{\bar{w}}_i \big( \beta E(\mathbf{s}_i) + \ln \hat{Z} \big), \qquad \mathbf{s}_i \sim q_{\theta},
\end{equation}
where the estimator of partition function $\hat Z$ is given by Eq.~(\ref{Zhat def}). There is qualitative difference between the estimator of the partition function and estimators of thermodynamical state functions, defined above: the $\hat Z$ estimator is unbiased, whereas taking logarithm introduces a systematic bias. However, as was shown in Ref.~\cite{PhysRevLett.126.032001}, such bias decreases as $\sim 1/N$ for large $N$ and can be neglected since the statistical uncertainty of $\ln \hat{Z}$ decreases as $\sim 1/\sqrt{N}$.

The variational estimates of $F$ and $S$ read \cite{2019PhRvL.122h0602W},
\begin{equation}
    F_{\textrm{HAN}} = \frac{1}{\beta} \frac{1}{N} \sum_{i=1}^N \big( \beta E(\mathbf{s}_i) + \ln \qb(\mathbf{s}_i) \big), \qquad \mathbf{s}_i \sim q_{\theta}
\end{equation}
and
\begin{equation}
    S_{\textrm{HAN}} = \frac{1}{N} \sum_{i=1}^N \ln \qb(\mathbf{s}_i), \qquad \mathbf{s}_i \sim q_{\theta}.
\end{equation}

\begin{figure*}
\begin{center}
\end{center}
\includegraphics[width=0.495\textwidth]{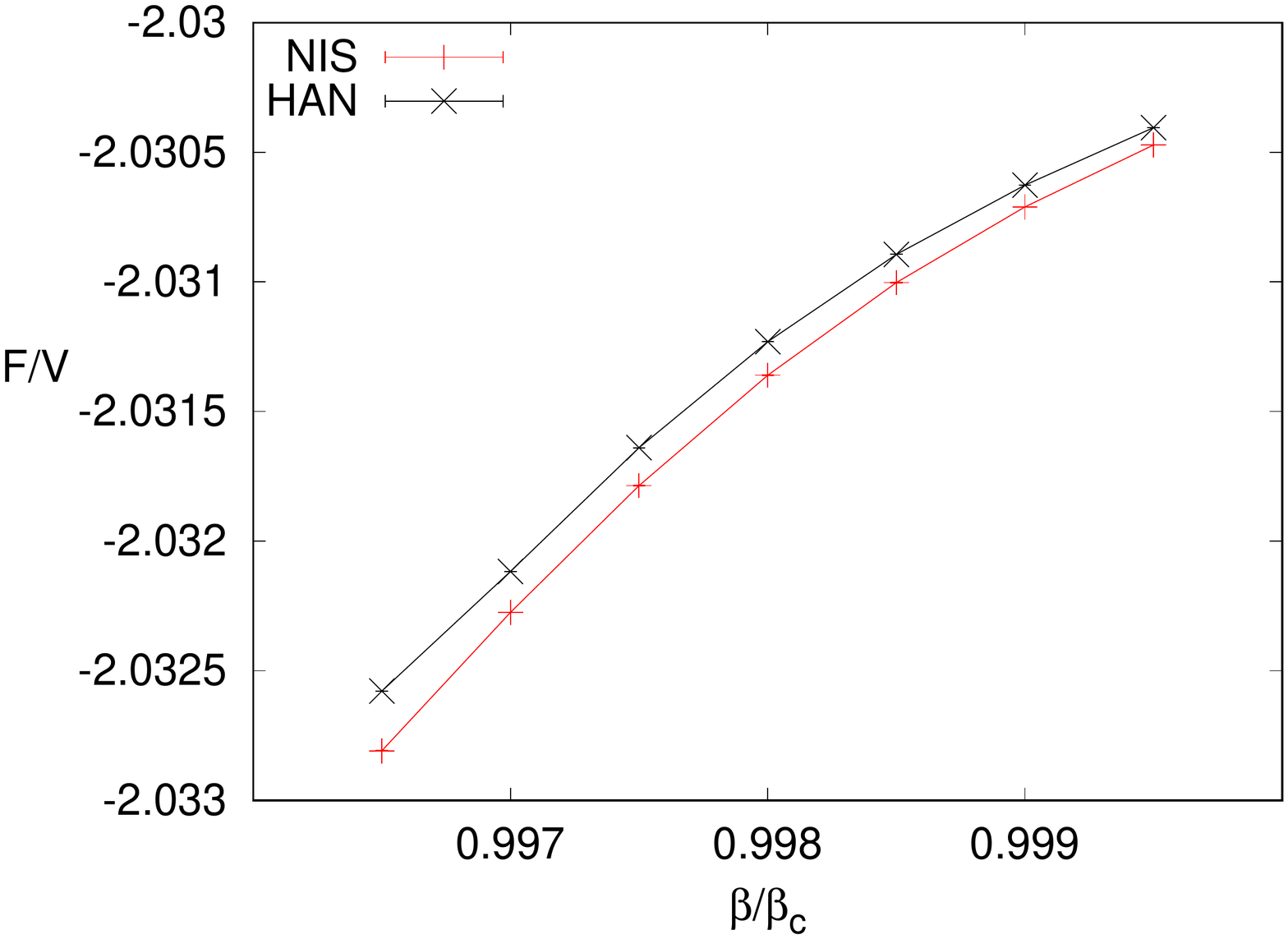}
\includegraphics[width=0.495\textwidth]{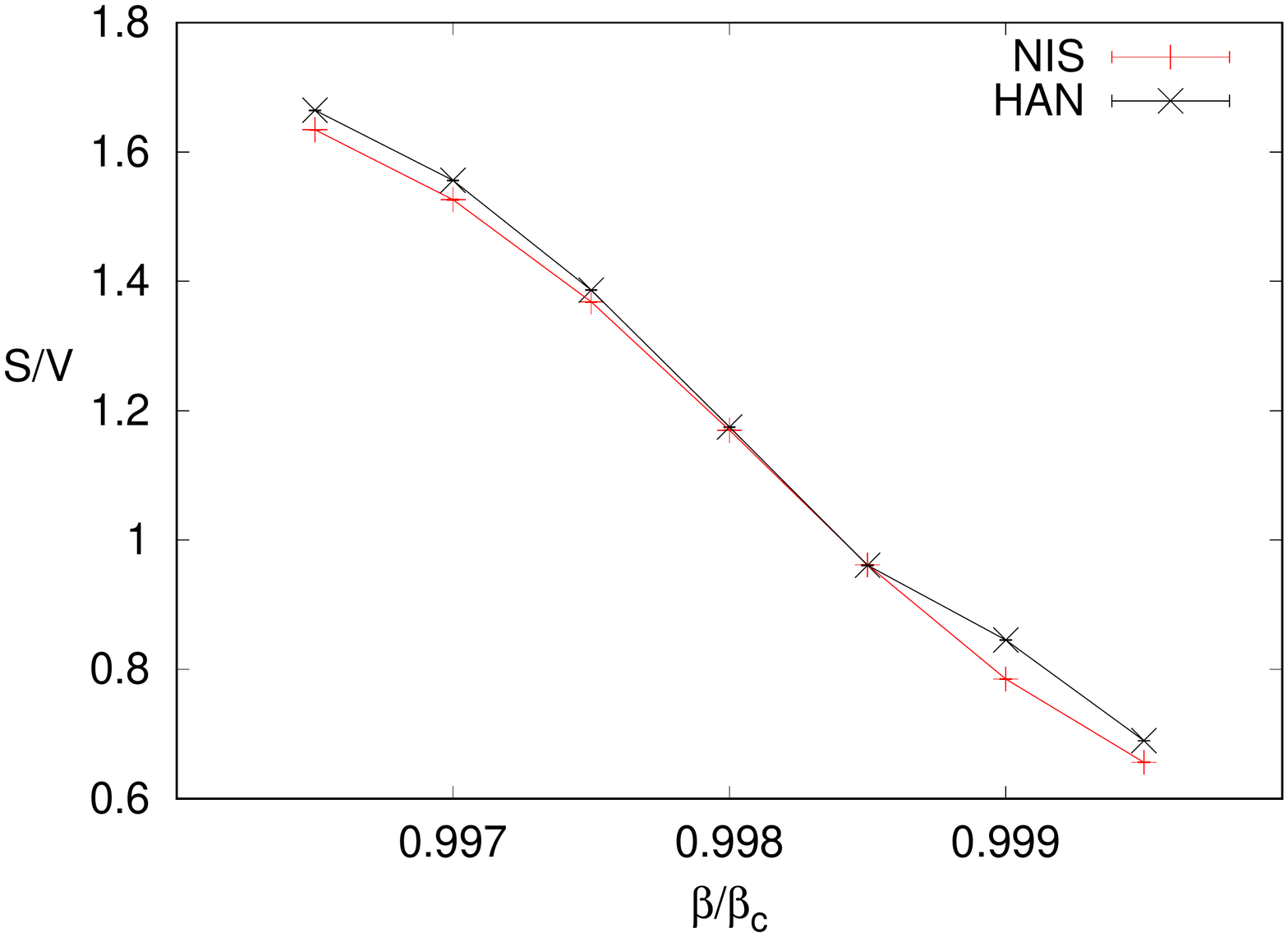}
\caption{Density of free energy (left panel) and entropy (right panel) for $L=32$ as a function of $\beta$. NIS data were obtained through reweighting to the Boltzmann probability distribution whereas HAN data correspond to the directly variationally approximated quantity from the HAN approach. Statistical errors are estimated through jackknife resampling and are smaller than the symbol size. For numerical data see table Tab.~\ref{tab. free energy}. \label{fig: free energy entropy}}
\end{figure*}

We have estimated $F$ and $S$ directly from $10^6$ configurations generated by the neural networks with the probability $\bar{q}_{\theta}$. We show the HAN results in Fig.~\ref{fig: free energy entropy} with black symbols. We also calculate $\hat{F}_{\textrm{NIS}}$ and $\hat{S}_{\textrm{NIS}}$ which are reweighted to the target probability distribution and correspond to exact results up to statistical uncertainties which we depict with red symbols. In both cases, free energy and entropy, the differences are small, however statistically significant. This again proves that $\bar{q}_{\theta}$ approximates $p$ quite closely. The statistical uncertainties have been estimated using  jackknife resampling. They are of the order of $10^{-7}$ for the free energy and $10^{-3}$ for the entropy, much smaller than the symbol size in Fig.~\ref{fig: free energy entropy}. The numerical values of $F_{\textrm{NIS}}$ and $S_{\textrm{NIS}}$ are also provided in Tab.~\ref{tab. free energy}. We stress that our approach based on generative neural network models allows calculating these quantities directly and independently at each value of the inverse temperature $\beta$, while most of the traditional Monte Carlo approaches are based on some form of integration of changes of thermodynamic observables along some line of changing $\beta$ (see Ref.~\cite{PhysRevLett.126.032001} for such analysis in $\phi^4$ field theory).

\begin{table}
\centering
\begin{tabular}{|c|l|c|}
\hline
$\beta/\beta_c$ & $F_{\textrm{NIS}}/V$ & $S_{\textrm{NIS}}/V$ \\
\hline
0.9965	&   -2.0328087(10)    &	1.63411(76)	    \\
0.9970  &   -2.03227453(60)	    &	1.52612(60)	    \\
0.9975  &   -2.03178624(60)	    &   1.36798(67)	    \\
0.9980	&   -2.03136002(63)     &	1.16976(73)     \\
0.9985	&   -2.03100296(65)     &	0.96147(75)	    \\
0.9990	&   -2.03071096(44)     &	0.78473(61)	    \\
0.9995  &	-2.03047153(42)     &	0.65613(53)	    \\
\hline
\end{tabular}
\caption{Numerical values of the free energy and entropy in the NIS approach and their statistical uncertainties. The corresponding plot is shown in Fig.~\ref{fig: free energy entropy}.}
\label{tab. free energy}
\end{table}

\section{Conclusions and discussion}
\label{sec. VII}

In this work, we have provided a numerical demonstration that the Neural Markov Chain Monte Carlo and Neural Importance Sampling provide a competitive approach to studying statistical systems in the vicinity of the first-order phase transition. We have described a hierarchical approach that uses a set of neural networks thanks to which we have considerably lowered the numerical cost of training and generating configurations. This allowed us to simulate the Potts model with $Q=12$ on a $32 \times 32$ lattice with statistical uncertainties more than one order of magnitude smaller than the ones obtained with the cluster algorithm on a set of configurations of the same size. In addition, our approach enables us to estimate thermodynamic observables such as free energy and entropy which we have calculated with high precision. We stress that these quantities are usually very difficult to obtain with traditional Markov Chain Monte Carlo methods. Hence, we believe that our results provide an encouraging example proving that the HAN approach may be helpful in understanding the thermodynamics of systems across first-order phase transitions.

As far as larger system extents are concerned, our experience with the cluster algorithm in the Potts model shows that it gets stuck in only one of the modes for the entire simulation of $10^6$ configurations at $L=64$ and $L=128$. This may be also inferred from the unfavorable scaling of the integrated autocorrelation times shown in Fig.~\ref{fig: wolff}. One may expect that with adequate training the NIS and NMCMC approaches might prove to be more ergodic. However, they require a more refined training strategy whose details are being currently developed.

\section*{Data open access}

Together with this manuscript, we publish our pytorch implementation of the HAN algorithm \cite{git}.
It is based on the implementation of VAN algorithm in the Ising model published together with Ref.~\cite{2019PhRvL.122h0602W}. We also publish the set of trained neural networks for the $Q=12$ Potts model at the seven temperatures. We believe that this will help to further improve the algorithm.

\section*{Acknowledgments}
Computer time allocation 'plgtmdlangevin2', 'plgnnformontecarlo' and 'plgng' on the Prometheus supercomputer hosted by AGH Cyfronet in Krak\'{o}w, Poland was used through the polish PLGRID consortium. T.S. kindly acknowledges support of the Polish National Science Center (NCN) Grants No.\,2019/32/C/ST2/00202 and 2021/43/D/ST2/03375. P.C. and P.K. acknowledge that this research was partially funded by the Priority Research Area Digiworld under the program Excellence Initiative – Research University at the Jagiellonian University in Kraków. P.K. and T.S thank Alberto Ramos and the University of Valencia for hospitality during the stay when this work was finalized.

\appendix

\section{Potts model with $Q=2$}
\label{ap. potts}

\begin{figure}
\begin{center}
  \includegraphics[width=0.495\textwidth]{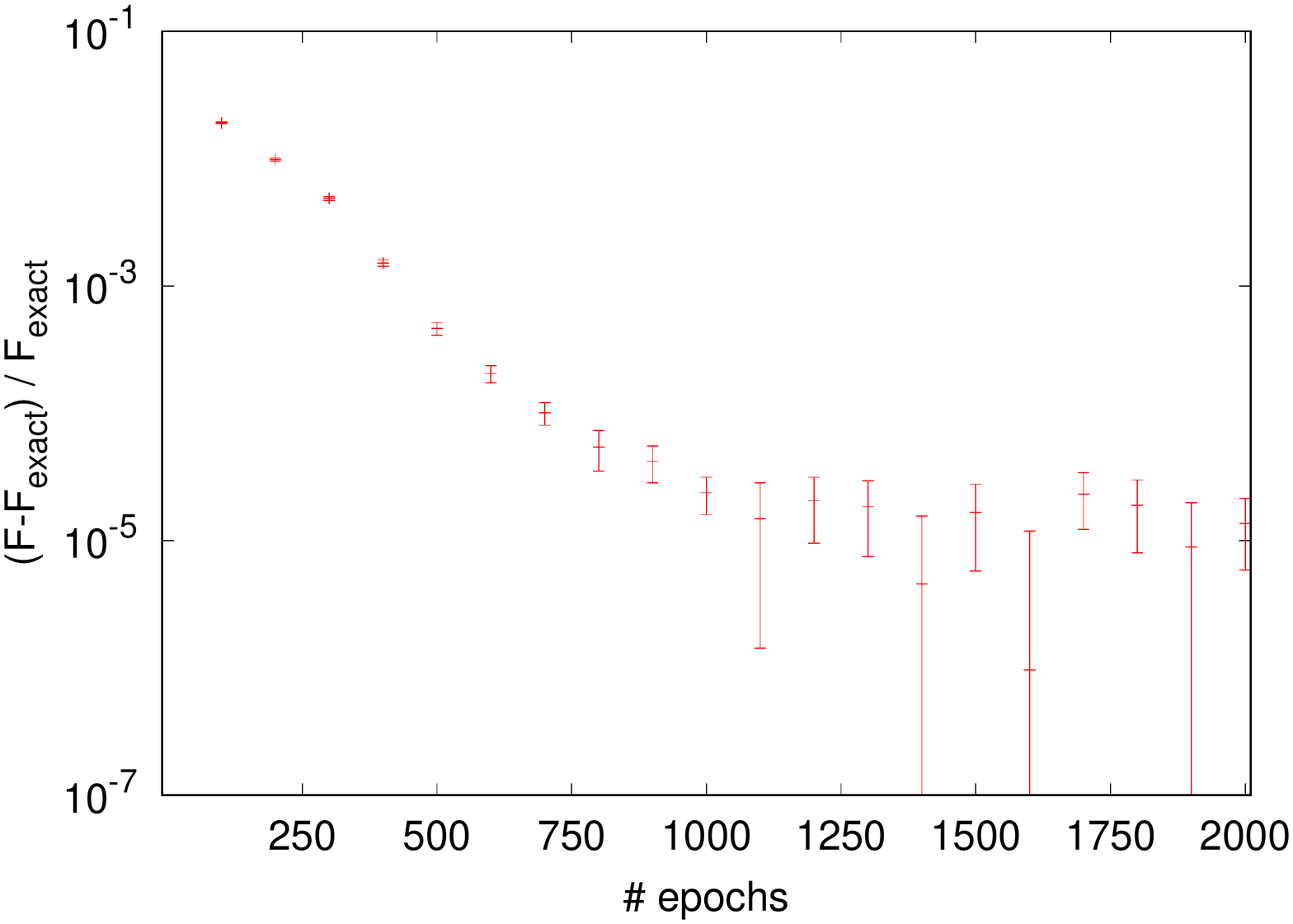} 
  \includegraphics[width=0.495\textwidth]{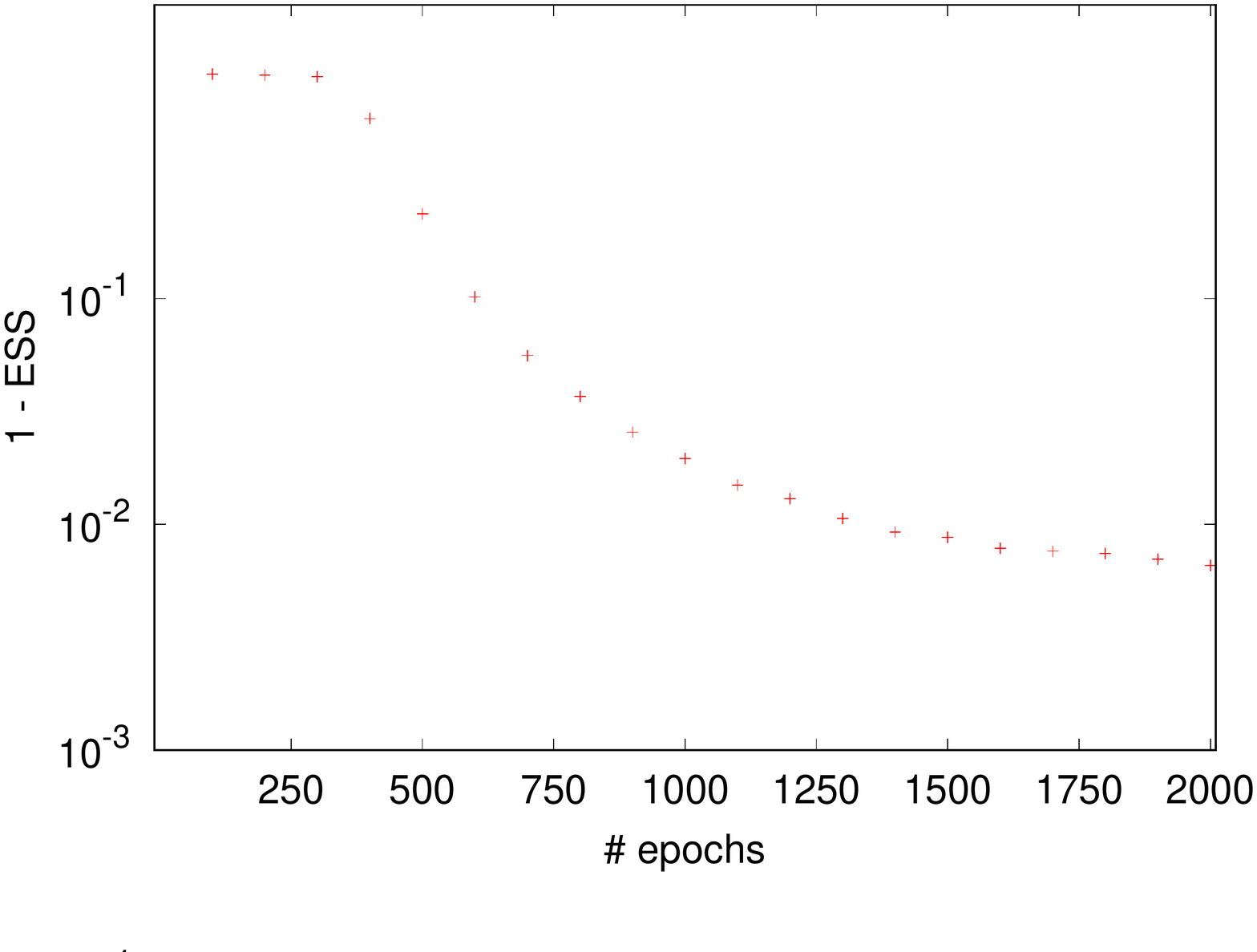} 
\end{center}
\caption{Free energy of the Ising model for $\beta=0.3$ and $L=16$ plotted versus the analytic value (upper panel) and (1-ESS) defined in Eq.~\eqref{ESS_definition}. Note that 2000 epochs are sufficient to train the hierarchical set of neural networks in this situation. \label{fig:training_quality}}
\end{figure}

Let us note that for $Q=2$ there exists an equivalence between the Potts model Eq.~\eqref{Potts_hamilt} and the exactly solvable Ising model, with the Hamiltonian
\begin{equation}
    H^I(\mathbf{s}) = - \sum_{\langle i,j \rangle} s^i \, s^j \qquad s^i, s^j \in \{-1,1\}.
\label{Ising_hamilt}
\end{equation}
The latter undergoes a second-order phase transition at $\beta_c = \frac{1}{2} \ln (1 + \sqrt{2}) \approx 0.441$ and the free energies of the two models are related as
\begin{equation}
    \frac{1}{2}\big(F^I(\beta)-2\big) = F^P(Q=2,2\beta),
\end{equation}
where by $F^I$ and $F^P$ we mean the free energies calculated for the Ising and Potts Hamiltonians, Eq.~\eqref{Ising_hamilt} and Eq.~\eqref{Potts_hamilt} respectively.
We use this relation to check the correctness of our algorithm by comparing our numerical outcomes with analytic results available for the two-dimensional Ising model on a square lattice of size $16\times 16$ in Fig.~\ref{fig:training_quality} (upper panel). We show on the logarithmic scale the difference between the analytic result and the outcome of HAN normalized by the analytic value. In the lower panel of the figure, we show 1-ESS as defined in Eq.~\eqref{ESS_definition} as a measure of the training quality. Note that in this simple situation the HAN approach provides satisfactory results after 2000 epochs of training only.

\section{Global symmetries}
\label{app:symm}

The gradient of the loss function Eq.~\eqref{eq:KL_loss_sym} can be approximated by (see also Ref.~\cite{2019PhRvL.122h0602W}),
\begin{equation}\label{eq:dkl_grad}
    \frac{1}{M}\sum_{i=1}^M \diff{\ln \qb(\b s^i)}{\b\theta}\left(\ln \qb(\b s^i)-\ln p(\b s^i)\right),
\end{equation}
assuming that configurations $\b s^i$ are sampled from distribution $\qb$.  However, it is correct also when they are sampled from $\q(\b s)$ as it is in our case. 

To see that let's modify our sampling procedure a little bit. Instead of sampling configuration $\v s$ directly from $\q$, we will first sample configuration $\v s'$ from $\q$  and then choose one $h_i$ at random with probability $M^{-1}$ and set  $\v s = h_i(\v s')$. 
The probability of sampling $\v s$ is equal to 
\begin{equation}
\frac{1}{M}\sum_{i} \q(h_i^{-1}(\v s))   =  \frac{1}{M}\sum_{i} \q(h_i(\v s)) = \qb(\v s). 
\end{equation}
For this equality to hold it is necessary that elements $h_i$ form a group $H$ (a subgroup of the symmetry group).  We can then use the formula \eqref{eq:dkl_grad} to estimate the gradient. 
However, because this formula is invariant under the action of each $h_i$ it does not matter if we use $h_i(\v s)$ or $\v s$ and the formula is correct even if we do not  transform the configuration.

The same reasoning can be applied to neural importance sampling (NIS) or neural Markov chain Monte-Carlo (NMCMC). If we transform configuration at random after sampling  we effectively sample from $\bar{q}_{\theta}$. But if our observables  are invariant under the action of $H$ then the actual transformation is unnecessary.

\section{Numerical details}
\label{ap. quality}

\subsection{Wolff's cluster algorithm}

The cluster algorithm was implemented as a single-threaded CPU program. All simulations were performed twice, once starting from a hot and once starting from a cold configuration. The thermalization period defined as the part until the magnetization evaluated in the two simulations converged to approximately the same value was rejected. Afterward, statistics of $10^6$ configurations, separated by one sweep, was generated. It took approximately 8h of running on an Intel(R) Core(TM) i5-7200U CPU @ 2.50GHz processor.

\subsection{HAN}

We used NVIDIA V100 GPGPU accelerators for our simulations. It took $\sim$ 8h to generate an ensemble of $10^6$ configurations for each investigated inverse temperature. For the training, we used approximately 130h of wall-clock time for each inverse temperature corresponding to approximately 250000 epochs. Inverse temperature $\beta$ was kept fixed during the training. We used smaller neural networks pretrained on systems $8 \times 8$ and $16 \times 16$ beforehand. The learning rate was adjusted during training, usually, it decreased as the training proceeded. We used the Adam optimizer, starting with a learning rate of order 0.001 and reaching $10^{-6}$ at the end of training.

\section{Improved partitioning}
\label{app. partitioning}

We implement one improvement compared to the algorithm described in Ref.~\cite{Bialas:2022qbs}, namely the largest neural network for the boundary spins is divided into two parts: the exterior boundary marked in red in the left panel of Fig.\ref{fig:sketch} and the interior which has a shape typical to the interior spins in the hierarchical algorithm. Hence, we exchange the original neural network of size $Q(4L-4)$ which had $\sim 16Q^2L^2$ parameters by two networks, one of size $Q(2L-1)$ with $\sim 4Q^2L^2$ parameters and another one with $Q(6L-5)$ input neurons and $Q(2L-1)$ output neurons and hence $\sim 12Q^2L^2$ parameters. Although the number of parameters is similar to the one in the original proposal, however, our experiences show that it is easier to train multiple smaller neural networks instead of one large. Moreover, the neural network for the interior spins can be re-used for simulations of larger systems, which was not possible in the previous approach.

\section{NIS estimate of energy density  }
\label{app:energy-density}
\newcommand{\delt}{\delta_{H(\v s ),E}}

\begin{figure}
    \centering
    \includegraphics[width=0.475\textwidth]{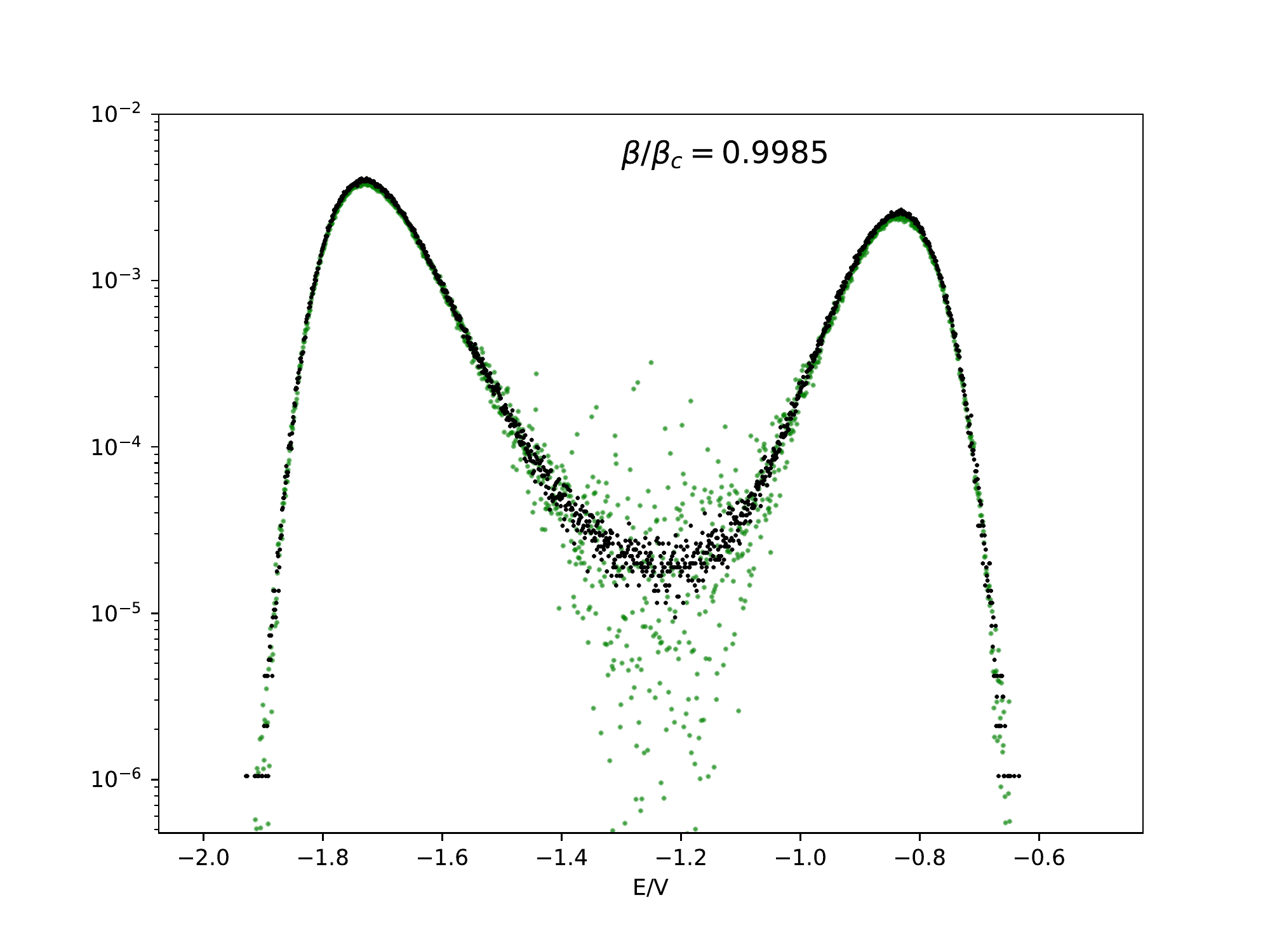}
    \caption{Energy density histogram estimated using NIS (green dots). Black dots are the results from Wolff cluster algorithm}
    \label{fig:nis-energy-density}
\end{figure}

In this section we will derive the NIS estimator of energy density. Similar argument applies to NMCMC, justifying our reasoning given in Section~\ref{sec. V}.

The energy density is given by:
\begin{equation}
    p(E) = \frac{Z(E)}{Z},
\end{equation}
where 
\begin{equation}\label{eq:Ze1}
    Z(E) = \sum_{\v s} e^{-\beta H(\v s)}\delt=\avg{\whb(\v s) \delt}_{\qb(\v s)}.
\end{equation}

Estimator for $Z(E)$ is given by:
\begin{equation}\label{eq:Ze-estimator1}
    \hat Z(E)=\frac{1}{N}\sum_{i=1}^N \whb(\v s_i)\delt
\end{equation}
and for $p(E)$ by:
\begin{equation}\label{eq:pe-estimator1}
    \hat{p}(E)=\frac{\sum_{i=1}^N \whb(\v s_i)\delt}{\sum_{i=1}^N \whb(\v s_i)}.
\end{equation}
The results for $\hat p(E)$ are presented in the Figure~\ref{fig:nis-energy-density} as green dots and compared with cluster algorithm (black dots).

In order to understand large fluctuation around $E/V \approx -1.2$ we calculate the variance of estimator \eqref{eq:Ze-estimator1}:
\begin{equation} 
   \var{\hat Z(E)}=\avg{\left(\hat Z(E)-Z(E) \right)^2}_{\qb(\v s)}.
\label{eq:Ze-variance1}
\end{equation}
After some straightforward calculations we obtain 
\begin{equation} 
   \var{\hat Z(E)}=
   \frac{Z^2}{N}\left(
p(E)w(E)-p(E)^2
\right),
\label{eq:Ze-variance_result}
\end{equation}
where 
\begin{equation}
    w(E)\equiv \left(\sum_{\v s: H(\v s)=E} 1 \right)^{-1} \sum_{\v s:H(\v s)=E}\frac{p(\v s)}{\qb(\v s)}.
\end{equation}
Note that the sum in the bracket is number of states with energy $E$. The quantity $w(E)$ measures how much on average $p(\v s)$ is greater than $\qb(\v s)$ for states with energy $E$. 

Because $Z$ is estimated from a much larger number of samples that each individual $p(E)$, its contribution to the variance of $p(E)$ is much smaller then \eqref{eq:Ze-variance1}. Therefore, we can write the relative error of $p(E)$ as:
\begin{equation}\label{eq:pe-rel-variance1}
   \frac{\sqrt{\var{\hat p(E)}}}{p(E)}\approx \frac{1}{\sqrt{N}}\sqrt{\frac{w(E)-p(E)}{p(E)}}.
\end{equation}

From this formula it follows that 
if for some energy $E$ the $\qb(\v s)$'s are smaller than $p(\v s)$'s by average factor $w(E)>1$, then the energy density error  will be roughly $\sqrt{w(E)}$ times larger than for perfectly trained network.

\bibliographystyle{ieeetr}
\bibliography{references2}

\begin{thebibliography}{10}

\bibitem{potts}
R.~B. {Potts} and C.~{Domb}, ``{Some generalized order-disorder
  transformations},'' {\em Proceedings of the Cambridge Philosophical Society},
  vol.~48, p.~106, Jan. 1952.

\bibitem{Gattringer:2010zz}
C.~Gattringer and C.~B. Lang, {\em {Quantum chromodynamics on the lattice}},
  vol.~788.
\newblock Berlin: Springer, 2010.

\bibitem{PhysRevLett.57.2607}
R.~H. Swendsen and J.-S. Wang, ``Replica monte carlo simulation of
  spin-glasses,'' {\em Phys. Rev. Lett.}, vol.~57, pp.~2607--2609, Nov 1986.

\bibitem{parallel_tempering}
D.~J. Earl and M.~W. Deem, ``Parallel tempering: Theory{,} applications{,} and
  new perspectives,'' {\em Phys. Chem. Chem. Phys.}, vol.~7, pp.~3910--3916,
  2005.

\bibitem{Albandea:2021lvl}
D.~Albandea, P.~Hern\'andez, A.~Ramos, and F.~Romero-L\'opez, ``{Topological
  sampling through windings},'' {\em Eur. Phys. J. C}, vol.~81, no.~10, p.~873,
  2021.

\bibitem{Albandea:2021kwe}
D.~Albandea, P.~Hern\'andez, A.~Ramos, and F.~Romero-L\'opez, ``{Improved
  topological sampling for lattice gauge theories},'' {\em PoS},
  vol.~LATTICE2021, p.~183, 2022.

\bibitem{DelDebbio:2021qwf}
L.~Del~Debbio, J.~M. Rossney, and M.~Wilson, ``{Efficient modeling of
  trivializing maps for lattice \ensuremath{\phi}4 theory using normalizing
  flows: A first look at scalability},'' {\em Phys. Rev. D}, vol.~104, no.~9,
  p.~094507, 2021.

\bibitem{DelDebbio:2021mts}
L.~Del~Debbio, J.~M. Rossney, and M.~Wilson, ``{Machine Learning Trivializing
  Maps: A First Step Towards Understanding How Flow-Based Samplers Scale Up},''
  {\em PoS}, vol.~LATTICE2021, p.~059, 2022.

\bibitem{Albandea:2022fky}
D.~Albandea, L.~Del~Debbio, P.~Hern\'andez, R.~Kenway, J.~M. Rossney, and
  A.~Ramos~Martinez, ``{Learning trivializing flows},'' {\em PoS},
  vol.~LATTICE2022, p.~001, 2023.

\bibitem{Albandea:2023wgd}
D.~Albandea, L.~Del~Debbio, P.~Hern\'andez, R.~Kenway, J.~M. Rossney, and
  A.~Ramos, ``{Learning Trivializing Flows},'' 2 2023.

\bibitem{Luscher:2009eq}
M.~Luscher, ``{Trivializing maps, the Wilson flow and the HMC algorithm},''
  {\em Commun. Math. Phys.}, vol.~293, pp.~899--919, 2010.

\bibitem{2019PhRvL.122h0602W}
D.~{Wu}, L.~{Wang}, and P.~{Zhang}, ``{Solving Statistical Mechanics Using
  Variational Autoregressive Networks},'' {\em Phys. Rev. Lett.}, vol.~122,
  p.~080602, Mar. 2019.

\bibitem{PhysRevD.100.034515}
M.~S. Albergo, G.~Kanwar, and P.~E. Shanahan, ``Flow-based generative models
  for markov chain monte carlo in lattice field theory,'' {\em Phys. Rev. D},
  vol.~100, p.~034515, Aug 2019.

\bibitem{2020PhRvE.101b3304N}
K.~A. {Nicoli}, S.~{Nakajima}, N.~{Strodthoff}, W.~{Samek}, K.-R. {M{\"u}ller},
  and P.~{Kessel}, ``{Asymptotically unbiased estimation of physical
  observables with neural samplers},'' {\em Phys. Rev. E}, vol.~101, p.~023304,
  Feb. 2020.

\bibitem{boltmann_generators_science}
F.~Noé, S.~Olsson, J.~Köhler, and H.~Wu, ``Boltzmann generators: Sampling
  equilibrium states of many-body systems with deep learning,'' {\em Science},
  vol.~365, no.~6457, p.~eaaw1147, 2019.

\bibitem{phiala}
M.~S. Albergo, D.~Boyda, D.~C. Hackett, G.~Kanwar, K.~Cranmer, S.~Racanière,
  D.~J. Rezende, and P.~E. Shanahan, ``Introduction to normalizing flows for
  lattice field theory,'' 2021.

\bibitem{Albergo:2021bna}
M.~S. Albergo, G.~Kanwar, S.~Racani\`ere, D.~J. Rezende, J.~M. Urban, D.~Boyda,
  K.~Cranmer, D.~C. Hackett, and P.~E. Shanahan, ``{Flow-based sampling for
  fermionic lattice field theories},'' {\em Phys. Rev. D}, vol.~104, no.~11,
  p.~114507, 2021.

\bibitem{Albergo:2022qfi}
M.~S. Albergo, D.~Boyda, K.~Cranmer, D.~C. Hackett, G.~Kanwar, S.~Racani\`ere,
  D.~J. Rezende, F.~Romero-L\'opez, P.~E. Shanahan, and J.~M. Urban,
  ``{Flow-based sampling in the lattice Schwinger model at criticality},'' {\em
  Phys. Rev. D}, vol.~106, no.~1, p.~014514, 2022.

\bibitem{Abbott:2022zhs}
R.~Abbott {\em et~al.}, ``{Gauge-equivariant flow models for sampling in
  lattice field theories with pseudofermions},'' {\em Phys. Rev. D}, vol.~106,
  no.~7, p.~074506, 2022.

\bibitem{Abbott:2022hkm}
R.~Abbott {\em et~al.}, ``{Sampling QCD field configurations with
  gauge-equivariant flow models},'' {\em PoS}, vol.~LATTICE2022, p.~036, 2023.

\bibitem{Hackett:2021idh}
D.~C. Hackett, C.-C. Hsieh, M.~S. Albergo, D.~Boyda, J.-W. Chen, K.-F. Chen,
  K.~Cranmer, G.~Kanwar, and P.~E. Shanahan, ``{Flow-based sampling for
  multimodal distributions in lattice field theory},'' 7 2021.

\bibitem{PhysRevLett.126.032001}
K.~A. Nicoli, C.~J. Anders, L.~Funcke, T.~Hartung, K.~Jansen, P.~Kessel,
  S.~Nakajima, and P.~Stornati, ``Estimation of thermodynamic observables in
  lattice field theories with deep generative models,'' {\em Phys. Rev. Lett.},
  vol.~126, p.~032001, Jan 2021.

\bibitem{Wang_2022}
L.~Wang, Y.~Jiang, L.~He, and K.~Zhou, ``Continuous-mixture autoregressive
  networks learning the kosterlitz-thouless transition,'' {\em Chinese Physics
  Letters}, vol.~39, p.~120502, dec 2022.

\bibitem{Bialas:2022qbs}
P.~Bia\l{}as, P.~Korcyl, and T.~Stebel, ``{Hierarchical autoregressive neural
  networks for statistical systems},'' {\em Comput. Phys. Commun.}, vol.~281,
  p.~108502, 2022.

\bibitem{WOLFF1989379}
U.~Wolff, ``Comparison between cluster monte carlo algorithms in the ising
  model,'' {\em Physics Letters B}, vol.~228, no.~3, pp.~379--382, 1989.

\bibitem{Baxter_1973}
R.~J. Baxter, ``Potts model at the critical temperature,'' {\em Journal of
  Physics C: Solid State Physics}, vol.~6, p.~L445, nov 1973.

\bibitem{Hammersley-Clifford}
P.~C. J.~M.~Hammersley, ``Markov fields on finite graphs and lattices,'' 1971.

\bibitem{Clifford90markovrandom}
P.~Clifford, ``Markov random fields in statistics,'' in {\em Disorder in
  Physical Systems. A Volume in Honour of John M. Hammersley}, Clarendon Press,
  1990.

\bibitem{Bialas:2021bei}
P.~Bia\l{}as, P.~Korcyl, and T.~Stebel, ``{Analysis of autocorrelation times in
  neural Markov chain Monte Carlo simulations},'' {\em Phys. Rev. E}, vol.~107,
  no.~1, p.~015303, 2023.

\bibitem{Liu}
J.~S. Liu, ``Metropolized independent sampling with comparisons to rejection
  sampling and importance sampling,'' {\em Statistics and Computing}, vol.~6,
  no.~2, pp.~113--119, 1996.

\bibitem{git}
\url{https://github.com/piotrkorcyl/hierarchical_autoregressive_networks_for_potts_model}.

\end{thebibliography}

\end{document}